\documentclass [12pt]{article}
\usepackage{amsmath,amssymb,cite}

\usepackage{feynmp}
\usepackage[vcentermath,enableskew]{youngtab}
\usepackage{tabularx}
\usepackage[english]{babel}
\usepackage[dvips]{graphicx}
\usepackage{indentfirst}
\usepackage{epsfig}
\numberwithin{equation}{section}

\begin{document}

\title{Matrix Model Fixed Point of Noncommutative Phi-Four}

\author{ Badis Ydri $^{a}$\footnote{Email:ydri@stp.dias.ie,~badis.ydri@univ-annaba.org.},  Rachid Ahmim$^{b}$ \\
$^{a}$ Institute of Physics, BM Annaba University,\\
BP 12, 23000, Annaba, Algeria.\\
$^{b}$  Department of Physics, El-Oued University,\\
BP 789, 39000, El-Oued, Algeria.
}

\maketitle
\abstract{In this article we exhibit explicitly the matrix model ($\theta=\infty$) fixed point of phi-four theory on noncommutative spacetime  with only two noncommuting directions   using the Wilson renormalization group recursion formula and the $1/N$ expansion of the zero dimensional reduction and then calculate the mass critical exponent $\nu$ and the anomalous dimension $\eta$ in various dimensions .}
\tableofcontents 
\section{Introduction and Summary of Results}
\subsection{Introduction}
The Wilson recursion formula is the oldest, most simple and most intuitive renormalization group approach which although approximate agrees very well with high temperature expansions \cite{Wilson:1973jj}. 

The goal in this article is to apply this method to the self-dual noncommutative phi-four in the matrix basis \cite{Langmann:2003if} which after appropriate non perturbative definition becomes an $N\times N$ matrix model where $N$  is a regulator in the noncommutative directions. More precisely we propose to employ, following \cite{Ferretti:1996tk,Ferretti:1995zn,Nishigaki:1996ts}, a combination of $i)$ the  Wilson  approximate renormalization group recursion formula and $ii)$ the solution to the corresponding  zero dimensional large $N$ counting problem given in our case by the hermitian Penner matrix model which can be turned into  a multi trace  hermitian quartic matrix model for large values of $\theta$. As discussed neatly in  \cite{Ferretti:1996tk} the virtue and power of combining these two methods lies in the crucial fact that all leading Feynman diagrams in $1/N$ will be counted correctly in this scheme including the so-called "setting sun" diagrams. As it turns out the recursion formula can also be integrated explicitly in the large $N$ limit which  in itself is a very desirable property.

In a previous work \cite{Ydri:2012nw} a non perturbative study of the Ising universality class fixed point in noncommutative $O(N)$ model was carried out using  precisely a combination of the above two methods. It was found that the Wilson-Fisher fixed point makes good sense only for sufficiently small values of $ \theta$ up to a certain maximal noncommutativity. In the current work we focus on the opposite limit of large $\theta$  although the $1/N$ expansion invoked in this article is different from the $1/N$ expansion of the $O(N)$ vector model since $N$ here has direct connection with noncommutativity itself. The central aim of this article as we will see is to exhibit as explicitly as possible the matrix model fixed point which describes the transition from the one-cut (disordered) phase to the two-cut (non-uniform ordered, stripe) phase in the same way that the Wilson-Fisher fixed point describes the transition from the disordered phase to the uniform ordered phase.

\subsection{Summary of Results}

We start by summarizing the main statements and results of this paper. 
We will be interested in  phi-four theory on a degenerate noncommutative Moyal-Weyl space with only two noncommuting coordinates ${\bf R}^{d}_{\theta}={\bf R}^D\times {\bf R}^2_{\theta}$ where $D=d-2$ with commutation relations $[\hat{x}_i,\hat{x}_j]=i\theta_{ij}$, $[\hat{x}_i,x_{\mu}]=[x_{\mu},x_{\nu}]=0$. The action takes the form 
 \begin{eqnarray}
S[\Phi]&=&\nu_2 \int d^{D}x~ Tr_{\cal H}\bigg[\hat{\Phi}^+\bigg(-\frac{1}{2}\hat{\partial}_i^2+\frac{1}{2}\Omega^2\tilde{X}_i^2-\frac{1}{2}{\partial}_{\mu}^2+\frac{{\mu}^2}{2}\bigg)\hat{\Phi}+\frac{\lambda}{4!}\hat{\Phi}^{+}\hat{\Phi}~\hat{\Phi}^{+}\hat{\Phi}\bigg].\nonumber\\
\end{eqnarray}
In the above equation  $\Omega=(B\theta)/2$, $\nu_2=\sqrt{{\rm det} (2\pi\theta})$ and $\tilde{X}_i=2(\theta^{-1})_{ij}X_j$ where $X_i=(\hat{x}_i+\hat{x}_i^R)/2$. This action is covariant under a duality transformation which exchanges among other things positions and momenta as $x_i\leftrightarrow \tilde{k}_i=B^{-1}_{ij}k_j$. The value ${\Omega}^2=1$ in particular gives an action which is invariant under this duality transformation.

By expanding the field in an appropriate basis (for example the Landau basis), introducing a cutoff $N$ in the noncommuting directions and a cutoff $\Lambda$ in the commuting directions and setting $\Omega^2=1$ we obtain the action

\begin{eqnarray}
S[M]&=&\int d^Dx Tr_N\bigg[\frac{1}{2}\partial_{\mu}M^+\partial_{\mu}M+\frac{1}{2}\mu^2M^+M+\frac{1}{2}r^2E\{M,M^+\}+\frac{u}{N}(M^+M)^2\bigg].\nonumber\\
\end{eqnarray}
\begin{eqnarray}
r^2=\frac{8\pi }{\nu_2}~,~u=\frac{\lambda}{4!}\frac{N}{\nu_2}~,~
E_{lm}=(l-\frac{1}{2}){\delta}_{lm}.
\end{eqnarray}
We will consider in the remainder only the case of hermitian matrices, viz
\begin{eqnarray}
 M=M^+. 
\end{eqnarray}
There are three independent parameters in this theory. These are the usual mass parameter $\mu^2$ and the quartic coupling constant $u$ plus the inverse noncommutativity $r^2=4/\theta$. The free propagator of this theory is simple given by
\begin{eqnarray}
\Delta_{mn}(p)=\frac{1}{p^2+\mu^2+r^2(m+n-1)}.
\end{eqnarray}
In the limit $\theta\longrightarrow \infty$ this propagator behaves as $1/(p^2+\mu^2)$. More precisely we have in this limit the useful properties

\begin{eqnarray}
 \sum_m{\Delta}^{r_1}_{mj_1}(p_1){\Delta}^{r_2}_{mj_2}(p_2)... \longrightarrow N{\Delta}^{r_1}_{n_0j_1}(p_1){\Delta}^{r_2}_{n_0j_2}(p_2)....
\end{eqnarray}
The Wilson renormalization group approach consists in general in the three main steps: $1)$ Integration, $2)$ Rescaling and $3)$ Normalization. In our case here we will supplement the first step of integration with two approximations $a)$ Truncation and $b)$ Wilson Recursion formula. 

We start by decomposing  the $N\times N$ matrix $M$ into an $N\times N$ background matrix $\tilde{M}$ and an $N\times N$ fluctuation matrix $m$, viz $M=\tilde{M}+m$. The background $\tilde{M}$ contains slow modes, i.e. modes with momenta less or equal than $\rho\Lambda$ while the fluctuation $m$  contains fast modes, i.e. modes with momenta  larger  than $\rho\Lambda$ where $0<\rho<1$. The integration step involves performing the path integral over the fluctuation $m$ to obtain an effective path integral over the background $\tilde{M}$ alone. We find

\begin{eqnarray}
Z
&=&\int d\tilde {M}e^{-S[\tilde{M}]}~e^{-\Delta S(\tilde{M})}.
\end{eqnarray}
The main goal is to compute the effective action $\Delta S(\tilde{M})$ which contains corrections to the operators already present in the original action $S[\tilde{M}]$ together with all possible effective interactions generated by the integration process. An exact formula for $\Delta S(\tilde{M})$ up to the fourth power in the field $\tilde{M}$ is given by the cumulant expansion (\ref{mainresult}).

The formula (\ref{mainresult}) is still very complicated. To simplify it and to get explicit equations we employ the so-called Wilson truncation and Wilson recursion formula. This is usually thought of as part of the integration step. Wilson truncation means that we calculate quantum corrections to only those terms which appear in the starting action. Wilson recursion formula is completely equivalent to the use in perturbation theory of the Polyakov-Wilson rules   given by the following two approximations: $1)$ We replace every internal propagator  $1/(k^2+\mu^2)$ by $1/(c^2+\mu^2)$ where $c$ is a constant taken to be equal to the cutoff $\Lambda$ and $2)$ We replace every momentum loop integral $\int_{\rho\Lambda}^{\Lambda} d^Dk/(2\pi)^D$ by another constant $v_D=\Lambda^D\hat{v}_D$ where the definition of $\hat{v}_D$ is obvious. This is a very long and tedious calculation. The end result is given by the sum of the three equations (\ref{r1}), (\ref{r2}) and (\ref{r3}).

By performing the second step of the Wilson renormalization group approach, i.e. by scaling momenta as $p\longrightarrow p/\rho$ so that the cutoff returns to its original value $\Lambda$ and the third and final step of the Wilson renormalization group approach consisting in rescaling the field in such a way that  the kinetic term is brought to its canonical form we obtain the effective action (\ref{effectiveactionSD2}). In position space this effective action takes the form

\begin{eqnarray}
 S+\Delta S&=&\frac{1}{2}\int d^Dx Tr_N(\partial_{\mu}\tilde{M}^{'})^2+\frac{\mu^{'2}}{2}\int d^Dx Tr_N\tilde{M}^{'2}+r^{'2}\int d^Dx Tr_NE \tilde{M}^{'2}\nonumber\\
&+&\frac{u^{'}}{N}\int d^Dx Tr_N \tilde{M}^{'4}.
\end{eqnarray}
The renormalized field $\tilde{M}^{'}$ is related to the bare field $\tilde{M}$ as follows. If $\bar{M}$ and $\bar{M}^{'}$  are the Fourier transforms of $\tilde{M}$ and $\tilde{M}^{'}$ respectively then
 \begin{eqnarray}
\bar{M}^{'}(p)=\rho^{\frac{2+D}{2}}\sqrt{Z(g,\mu^2)+\frac{r^2N}{(c^2+\mu^2)^2}\Delta Z(g,\mu^2)}\bar{M}(\rho p).\label{wave}
\end{eqnarray}
The renormalized mass ${\mu}^{'2}$, the renormalized quartic coupling constant ${u}^{'}$ and the renormalized  inverse noncommutativity $r^{'2}$ are given by 
\begin{eqnarray}
{\mu}^{'2}=\frac{\rho^{-2}}{Z(g,\mu^2)}\bigg[\Gamma(g,\mu^2)+r^2 N\Delta \Gamma(g)-\frac{r^2N}{(c^2+\mu^2)^2}\frac{\Gamma(g,\mu^2)\Delta Z(g,\mu^2)}{Z(g,\mu^2)}\bigg].\label{rgI1}
\end{eqnarray}
\begin{eqnarray}
r^{'2}=\frac{\rho^{-2}r^2}{Z(g,\mu^2)}\bigg[\Gamma_e(g)+\frac{r^2(N+1)}{c^2+\mu^2}\Delta \Gamma_e(g)-\frac{r^2N}{(c^2+\mu^2)^2}\frac{\Gamma_e(g)\Delta Z(g,\mu^2)}{Z(g,\mu^2)}\bigg].\label{rgI2}
\end{eqnarray}
\begin{eqnarray}
 {u}^{'}=u\frac{\rho^{-\epsilon}}{Z^2(g,\mu^2)}\frac{1}{4g}\bigg[{\Gamma}_4(g)+\frac{r^2N}{c^2+\mu^2}\Delta\Gamma_4(g)-\frac{2r^2N}{(c^2+\mu^2)^2}\frac{\Gamma_4(g)\Delta Z(g,\mu^2)}{Z(g,\mu^2)}\bigg]. \label{rgI3}
\end{eqnarray}
The effective coupling $g$ is defined by 

\begin{eqnarray}
 g=\frac{v_Du}{(c^2+\mu^2+r^2N)^2}.
\end{eqnarray}
The various functions $Z(g,\mu^2)$, $\Gamma(g,\mu^2)$, $\Gamma_2(g)$ and $\Gamma_4(g)$ are known non perturbatively whereas we were able to determine the functions $\Delta Z(g,\mu^2)$, $\Delta \Gamma(g)$, $\Delta \Gamma_4(g)$, $\Delta \Gamma_e(g)$ and $\Gamma_e(g)$ only perturbatively.  These functions are summarized in table (\ref{table}). 

The process which led from the bare coupling constants  $\mu^2$, $r^2$ and $u$ to the renormalized coupling constants ${\mu}^{'2}$, $r^{'2}$ and $u^{'}$ can be repeated an arbitrary number of times. The bare coupling constants will be denoted by $\mu_{0}^{2}$, $r_{0}^{2}$ and $u_{0}$ whereas the the renormalized coupling constants  at the first step of the renormalization group procedure will be denoted by  $\mu_{1}^{2}$, $r_{1}^{2}$ and $u_{1}$. At a generic step $l+1$ of the renormalization group process the renormalized coupling constants  $\mu_{l+1}^{2}$, $r_{l+1}^{2}$ and $u_{l+1}$ are related to their previous values $\mu_{l}^{2}$, $r_{l}^{2}$ and $u_{l}$ by precisely the above renormalization group equations.  The effective coupling constant $g_l$ will of course be given in terms of  $\mu_{l}^{2}$, $r_{l}^{2}$ and $u_{l}$ by the same formula that related $g_0$ to $\mu_{0}^{2}$, $r_{0}^{2}$ and $u_{0}$. We are therefore interested in renormalization group flow in a $3-$dimensional parameter space generated by the mass $\mu^2$, the quartic coupling constant $u$ and the harmonic oscillator coupling constant (inverse noncommutativity) $r^2$.

We have reached the stage where it is very hard to push any further by pure analytical means and therefore we have to turn to numerical tools. In any case the renormalization group approach was originally devised with numerical approximations  in mind \cite{Wilson:1973jj}. See also  \cite{Bagnuls:2000ae}.

A renormalization group fixed point is a point in the space parameter which is invariant under the renormalization group flow. If we denote the fixed point by $\mu_*^2$, $r_{*}^2$ and $u_*$     then we must set $\mu=\mu^{'}=\mu_*$, $r=r^{'}=r_*$ and $u=u^{'}=u_*$ as well as $g=g^{'}=g_*$ in the above renormalization group equations (\ref{rgI1}), (\ref{rgI2}) and (\ref{rgI3}). The  matrix model fixed point corresponding to infinite noncommutativity is given by the following equations 
\begin{eqnarray}
{r}_{*}^2=0.
\end{eqnarray}
\begin{eqnarray}
f({g}_{*})=0.
\end{eqnarray}

\begin{eqnarray}
\hat{\mu}_*^2=\frac{\alpha_*}{1-\alpha_*}.
\end{eqnarray}

\begin{eqnarray}
\hat{u}_*=\frac{g_{*}(1+\hat{\mu}_*^2)^2}{\hat{v}_D}.
\end{eqnarray}
The functions $f$ and $\alpha$ are given by (with $\epsilon_0=\epsilon/D$ and $\epsilon=4-D$)
\begin{eqnarray}
f(g)&=&1+\frac{1}{\rho^2}- \sqrt{(1-\frac{1}{\rho^2}-2\epsilon_0 Z_2(g))^2+\frac{8}{\rho^2}(1+\epsilon_0)Z_2(g)(\Gamma_2(g)-1)}\nonumber\\
 &+&2\sqrt{\rho^{-\epsilon}\frac{{\Gamma}_4(g)}{4g}}-2Z_2(g)\epsilon_0.
\end{eqnarray}
\begin{eqnarray}
 \alpha(g)=\frac{1}{4(1+\epsilon_0)Z_2(g)}\bigg[-2- 2\sqrt{\rho^{-\epsilon}\frac{{\Gamma}_4(g)}{4g}}+4Z_2(g)\epsilon_0\bigg].
\end{eqnarray}
We can immediately see that this fixed point is fully determined by the functions $\Gamma_2$, $Z_2$ and $Z_4$ which are known non-perturbatively. These functions are the $2-$point proper vertex, the wave function renormalization and the $4-$point proper vertex respectively of the quartic matrix model. The results of this calculation are shown on table (\ref{table1}). In our approximation   we have checked that there is always a non-trivial fixed point for any value of $\rho$ in the interval $0<\rho<1$. 

We can also compute the mass critical exponent $\nu$ and the anomalous dimension $\eta$ within this scheme.
From the wave function renormalization (\ref{wave}) we compute immediately the anomalous dimension. We find

\begin{eqnarray}
  \eta
&=&\frac{\epsilon}{2}-\frac{\ln(\Gamma_4(g_*)/4g_{*})}{2\ln\rho}.
\end{eqnarray}
The computation of the mass critical exponent $\nu$ requires linearization of the renormalization group equations (\ref{rgI1}), (\ref{rgI2}) and (\ref{rgI3}). The linearized renormalization group equations are of the form (with $\delta G=G-G_*$ where $G=(G_1=\mu^2,G_2=u,G_3=r^2)$ )
 \begin{eqnarray}
\delta G^{'}={M}(G_*,\rho)\delta G^{}.
\end{eqnarray}
The matrix $M$ in our case is of the form
\begin{eqnarray}
\left( \begin{array}{ccc}
 M_{11} & M_{12} & M_{13}\\
 M_{21} & M_{22} & M_{23} \\
0 & 0 & M_{33}
\end{array} \right).
\end{eqnarray} 
We find that $\lambda_3=\rho^{-2}\Gamma_e(G_*)/{Z(G_*,\mu_*^2)}>1$ and hence $r^2$ is a relevant coupling constant like the mass. However the function $\Gamma_e(g)$ used in this formula is only known perturbatively and hence this conclusion should be taken with care. 

The two remaining eigenvalues are determined from the linearized renormalization group equations in the $2-$dimensional space generated by $G_1=\mu^2$ and $G_2=u$. As it turns out this problem depends only on functions which are fully known non perturbatively. The eigenvalues $\lambda_1(\rho)$ and $\lambda_2(\rho)$ can be determined from the trace and determinant of $M$ in an obvious way. It is not difficult to convince ourselves that these renormalization group eigenvalues must scale as
\begin{eqnarray}
\lambda_{\alpha}(\rho)=\lambda_{\alpha}(1)\rho^{-y_{\alpha}}.\label{beha1}
\end{eqnarray}
This formula (\ref{beha1}) was used as a crucial test for our numerical calculations. In particular we have determined by means of this formula the range of the dilatation parameter $\rho$ over which the logarithm  of the eigenvalues scale  linearly with $\ln \rho$. It is natural to expect this behavior to hold only if the renormalization group steps are sufficiently small so not to alter drastically the infrared physics of the problem. 



For $D=2$ the renormalization group steps can be thought of as small and the behavior (\ref{beha1}) holds in the regime $\ln \rho < -1$. As it happens this is the most important case corresponding to $d=4$. We find explicitly the following fits

\begin{eqnarray}
\ln |\lambda_1|=-1.296 \ln \rho +0.412~,~\ln |\lambda_2|=0.435 \ln \rho +2.438.
\end{eqnarray}
We conclude immediately that the scaling field $u_1$ corresponding to the mass  is relevant while the scaling field $u_2$ corresponding to the quartic coupling constant is irrelevant. This is the usual conclusion in $d=4$. The critical exponents in $D=2$ ($d=4$) are given respectively by
\begin{eqnarray}
y_1=1.296 ~,~y_2=-0.435.
\end{eqnarray}
The mass critical exponent is given by the inverse of the critical exponent $y_1$ associated with the relevant direction, viz $\nu=1/y_1$. The corresponding results for $y_1$ and $\nu$ are included in table (\ref{table3}).
We note that the results shown in table (\ref{table3}) are very close to the average value of $2/d$ and $2/D$, viz
\begin{eqnarray}
\nu=\frac{1}{d}+\frac{1}{D}.
\end{eqnarray}

\subsection{Outline} This article is organized as follows. In section $2$ we write down noncommutative phi-four in the matrix basis  and discuss some useful approximations involving  the propagator at the self-dual point which are valid for large $\theta$. In section $3$ we consider the dimensional reduction of noncommutative phi-four and some of its properties. In particular we will derive a very simple non perturbative equation for the $2-$point proper vertex which will be used to test the results obtained later using the $1/N$ expansion and the recursion formula.  
In section $4$ we perform the tedious task of deriving the renormalization group equations which control the flow of the coupling constants of the model (three in this case) using the Wilson renormalization group recursion formula and $1/N$ expansion. The most difficult piece of the calculation as we will see is wave function renormalization. In section $5$ we derive the non trivial fixed point and the associated critical exponents $\nu$ and $\eta$ by solving numerically  via the Newton-Raphson algorithm the renormalization group equations and discuss some of relevant physics. In section $6$ we extend the analysis to the Grosse-Vignes-Tourneret model which involves  an extra term, the double trace operator $(Tr_N M)^2$, required for the renormalizability of the theory. We conclude in section $7$ by a summary of the obtained results and brief outlook. We have also included three appendices for completeness and for the convenience of interested readers.

\section{The Noncommutative Phi-Four Theory}
\subsection{The Model}
Let us consider a phi-four theory on a generic noncommutative Moyal-Weyl space ${\bf R}^{d}_{\theta}$. We  introduce non-commutativity in momentum space by introducing a minimal coupling to a constant background  magnetic field $B_{ij}$ as was done originally by Langmann, Szabo and Zarembo in \cite{Langmann:2003if}. The most general action with a quartic potential takes in the operator basis the form 
\begin{eqnarray}
S=\sqrt{\det(2\pi{\theta})}Tr_{\cal H}\bigg[\hat{\Phi}^{+}\bigg(-\sigma \hat{D}_i^{2}-\tilde{\sigma}\hat{C}_i^{2}+\frac{{\mu}^{2}}{2}\bigg)\hat{\Phi}+\frac{\lambda}{4!}\hat{\Phi}^{+}\hat{\Phi}~\hat{\Phi}^{+}\hat{\Phi}+\frac{\lambda^{'}}{4!}\hat{\Phi}^{+}\hat{\Phi}^+~\hat{\Phi}\hat{\Phi}\bigg].\label{actionbasic}
\end{eqnarray}
In this equation $\hat{D}_i=\hat{\partial}_i-iB_{ij}X_j$ and $\hat{C}_i=\hat{\partial}_i+iB_{ij}X_j$ where $X_i=(\hat{x}_i+\hat{x}_i^R)/{2}$. In the original  Langmann-Szabo model we choose $\sigma=1$, $\tilde{\sigma}=0$ and $\lambda^{'}=0$ which as it turns out leads to a trivial model \cite{Langmann:2002cc}.
 
The famous Grosse-Wulkenhaar model corresponds to $\sigma=\tilde{\sigma}$ and $\lambda^{'}=0$. We choose without any loss of generality $\sigma=\tilde{\sigma}=1/4$. The Grosse-Wulkenhaar model corresponds to the addition of a harmonic oscillator potential to the kinetic action which modifies and thus allows us to control the IR behavior of the theory. A particular version of this theory was shown to be renormalizable by Grosse and Wulkenhaar in \cite{Grosse:2004yu}. The action of interest in terms of the star product is  given by 

\begin{eqnarray}
S&=&\int
d^dx~ \bigg[\Phi^+\bigg(-\frac{1}{2}{\partial}_i^2+\frac{1}{2}(B_{ij}x_j)^2+\frac{{\mu}^2}{2}\bigg)\Phi+\frac{\lambda}{4!}\Phi^+*\Phi*\Phi^+*\Phi\bigg].
\end{eqnarray}
Equivalently
\begin{eqnarray}
S
&=&\int d^{d}x \bigg[{\Phi}^+\bigg(-\frac{1}{2}{\partial}_i^2+\frac{1}{2}{\Omega}^2\tilde{x}_i^2+\frac{{\mu}^{2}}{2}\bigg){\Phi}+\frac{\lambda}{4!}{\Phi}^+*{\Phi}*{\Phi}^+*{\Phi}\bigg].
\end{eqnarray}
The harmonic oscillator coupling constant $\Omega$ is defined by $\Omega^2=B^2\theta^2/4$ whereas the coordinate $\tilde{x}_i$ is defined by $\tilde{x}_i=2({\theta}^{-1})_{ij}x_j$. It was shown in  \cite{Langmann:2002cc} that this action is covariant under a duality transformation which exchanges among other things positions and momenta as $x_i\leftrightarrow \tilde{k}_i=B^{-1}_{ij}k_j$. The value ${\Omega}^2=1$ in particular gives an action which is invariant under this duality transformation. The theory at  ${\Omega}^2=1$ is essentially the original Langmann-Szabo model.

Let us consider now a phi-four theory on a noncommutative Moyal-Weyl space with only two noncommuting coordinates, viz ${\bf R}^{d}_{\theta}={\bf R}^D\times {\bf R}^2_{\theta}$ where $D=d-2$. This is the degenerate case. The above action generalizes to

\begin{eqnarray}
S
&=&\int d^{d}x \bigg[{\Phi}^+\bigg(-\frac{1}{2}{\partial}_i^2+\frac{1}{2}{\Omega}^2\tilde{x}_i^2-\frac{1}{2}{\partial}_{\mu}^2+\frac{{\mu}^{2}}{2}\bigg){\Phi}+\frac{\lambda}{4!}{\Phi}^+*{\Phi}*{\Phi}^+*{\Phi}\bigg].
\end{eqnarray}
The index $i$ runs over the noncommuting directions while the index $\mu$ runs over the commuting directions.
Under the field/operator Weyl map we can rewrite the action as

\begin{eqnarray}
S&=&\nu_2 \int d^{D}x~ Tr_{\cal H}\bigg[\hat{\Phi}^+\bigg(-\frac{1}{2}\hat{\partial}_i^2+\frac{1}{2}\Omega^2\tilde{X}_i^2-\frac{1}{2}{\partial}_{\mu}^2+\frac{{\mu}^2}{2}\bigg)\hat{\Phi}+\frac{\lambda}{4!}\hat{\Phi}^{+}\hat{\Phi}~\hat{\Phi}^{+}\hat{\Phi}\bigg].\label{25}
\end{eqnarray}
The Planck volume $\nu_2$ is defined by $\nu_2=\sqrt{{\rm det} (2\pi\theta})$. It may not be obvious that the above action enjoys the same covariance under duality transformation as the action in the non-degenerate case. This is indeed true as we show explicitly in the first appendix. All the steps used in the non-degenerate case go virtually unchanged in the degenerate case and the noncommutative directions act merely as a spectator. 

It was established in \cite{Grosse:2008df} that the action (\ref{25}) is renormalizable if we add an appropriate  counterterm which was also determined. This agrees with our finding in this article where it will be shown using the Wilson renormalization group recursion formula that the action (\ref{25}) admits a non-trivial fixed point which therefore entails renormalizability \cite{ZinnJustin:2002ru}. In the Wilson renormalization group approach it is expected that the counterterm of \cite{Grosse:2008df} together with all possible effective interactions  will be generated in the process of integration and rescaling and as a consequence it will not be included  in (\ref{25}) in most of this article with the exception of section $6$. It remains however an interesting question in its own right to know what is precisely the effect of this counterterm if it is included.  In section $6$ the first steps towards this goal will be taken.
 
We can expand the scalar fields in the  Landau basis $\{\hat{\phi}_{m,n}\}$ as (with $x$ standing for  commuting coordinates)
\begin{eqnarray}
\hat{\Phi}=\frac{1}{\sqrt{\nu_2}}\sum_{m,n=1}^{\infty}M_{mn}(x)\hat{\phi}_{m,n}~,~\hat{\Phi}^{+}=\frac{1}{\sqrt{\nu_2}}\sum_{m,n=1}^{\infty}M_{mn}^{*}\hat{\phi}_{m,n}^{+}.
\end{eqnarray}
The Landau basis is constructed for example in \cite{GraciaBondia:1987kw}. The infinite dimensional matrix $M$ should be thought of as a compact operator acting on the separable Hilbert space ${H}={\cal S}({\bf N})$ of Schwartz sequences $(a_m)_{m\geq 1}$ with sufficiently rapid decrease as $m\longrightarrow \infty$.  In the Landau basis the action becomes

\begin{eqnarray}
S&=&\int d^Dx Tr_H\bigg[\frac{1}{2}\partial_{\mu}M^+\partial_{\mu}M+\frac{1}{2}\mu^2M^+M+\frac{1}{2}r^2E\{M,M^+\}+\frac{u}{N}(M^+M)^2+{\rm remainder}\bigg].\nonumber\\
\end{eqnarray}
The coupling constants $r^2$ and $u$ are defined by 
\begin{eqnarray}
r^2=\frac{4\pi (\Omega^2+1)}{\nu_2}~,~u=\frac{\lambda}{4!}\frac{N}{\nu_2}.
\end{eqnarray}
The remainder is given by
\begin{eqnarray}
{\rm remainder}=\frac{1}{2}r^2\sqrt{\omega}(\Gamma^+M^+\Gamma M+M^+\Gamma^+M\Gamma)~,~\sqrt{\omega}=\frac{1-\Omega^2}{1+\Omega^2}.
\end{eqnarray}
The matrices $\Gamma$ and $E$ are given by
\begin{eqnarray}
(\Gamma)_{lm}=\sqrt{m-1}{\delta}_{lm-1}~,~(E)_{lm}=(l-\frac{1}{2}){\delta}_{lm}.
\end{eqnarray}
We regularize the theory by taking $M$ to be an $N\times N$ matrix. We will also restrict the momenta along the commuting directions to be less or equal than a hard cutoff $\Lambda$.  The states ${\phi}_{l,m}(x)$ with  $l,m < N$ where $N$ is some large integer correspond to a cut-off in position and momentum spaces \cite{Grosse:2004yu}. The infrared cut-off is found to be proportional to $R=\sqrt{2\theta N}$ while the UV cut-off is found to be proportional to $\Lambda_N=\sqrt{8N/\theta}$. In \cite{Langmann:2003if} a double scaling strong noncommutativity limit  in which $N/\theta$ (and thus $\Lambda_N$)  is kept fixed was considered. In our analysis here we will take a different strong noncommutativity limit in which we take $\theta\longrightarrow \infty$ first and then  $N\longrightarrow \infty$. The trace $Tr_{H}$ will be replaced by $Tr_N$ with $Tr_N 1=N$.

This is a deformed principal chiral model. Indeed if we set the remainder equal to zero and choose $\mu^2=-4u/N$ we see immediately that in the limit $u\longrightarrow \infty$ the partition function is localized around configurations satisfying $M^+M=1$ with an action given by $(\int d^Dx~Tr_N\partial_{\mu}M^+\partial_{\mu}M)/2$. This is the usual definition of a principal chiral field in dimension $D=d-2$. The potential term provides a smoother definition of the model while the remainder gives a deformation due to non-commutativity. 

\subsection{The Propagator}
The free propagator simplifies in the limit $\theta\longrightarrow \infty$. Indeed in this limit the above action reduces to a $U(N)$ matrix model in $D=d-2$ dimensions with propagator given by $1/(p^2+\mu^2)$ \cite{Becchi:2003dg}. A more interesting limit is $\Omega^2\longrightarrow 1$. This is the self-dual theory in which the off diagonal quadratic terms vanish, i.e. the remainder is $0$. This theory was studied extensively from other perspective in \cite{Disertori:2006nq} where it was shown that the beta function vanishes to all orders. The action in this case is given by
\begin{eqnarray}
S&=&\int d^Dx Tr_N\bigg[\frac{1}{2}\partial_{\mu}M^+\partial_{\mu}M+\frac{1}{2}\mu^2M^+M+\frac{1}{2}r^2E\{M,M^+\}+\frac{u}{N}(M^+M)^2\bigg].\label{model}
\end{eqnarray}
We will consider in the remainder only the case of hermitian matrices, viz
\begin{eqnarray}
M=M^+.
\end{eqnarray}
There are three independent parameters in this theory. These are the usual mass parameter $\mu^2$ and the quartic coupling constant $u$ plus the inverse noncommutativity $r^2=4/\theta$. The free propagator of this theory is also simple given by
\begin{eqnarray}
<m_{nm}(x)m_{lk}(y)>_0
&=&\delta_{n,k}\delta_{m,l}\int \frac{d^Dp}{(2\pi)^D}\Delta_{mn}(p) e^{ip(x-y)}.
\end{eqnarray}
\begin{eqnarray}
\Delta_{mn}(p)=\frac{1}{p^2+\mu^2+r^2(m+n-1)}.\label{appr0}
\end{eqnarray}
In the next section we will need extensively the sum $\sum_m\Delta_{mn}(p)$. For $\theta=\infty$ this sum is obviously of order $N$. Including also the subleading $1/\theta$ correction this sum  takes then the form
\begin{eqnarray}
\sum_{m}{\Delta}_{mj}(p)\longrightarrow N {\Delta}_{n_0j}(p)~,~n_0=\frac{N+1}{2}.\label{appr1}
\end{eqnarray}
A straightforward generalization of (\ref{appr1}) is 

\begin{eqnarray}
 \sum_m{\Delta}^{r_1}_{mj_1}(p_1){\Delta}^{r_2}_{mj_2}(p_2)... \longrightarrow N{\Delta}^{r_1}_{n_0j_1}(p_1){\Delta}^{r_2}_{n_0j_2}(p_2)....\label{appr2}
\end{eqnarray}
The structure of the $1/N$ expansion is quite complicated for generic values of $\theta$. However, in the limit in which we take $\theta\longrightarrow \infty$ first and then $N\longrightarrow \infty$ we find that the $1/N$ expansion becomes manageable.  More precisely if  we  assume that $\theta$ is sufficiently large so that  we are allowed to use the approximations (\ref{appr1}) and (\ref{appr2})  
 then the Feynman diagrams that will dominate the large $N$ limit are exactly those of the hermitian matrix model in $D$ dimensions.

\section{ Quartic and Penner Matrix Models}
\subsection{The Disordered-to-Non-Uniform-Ordered (or Matrix) Transition}
As mentioned on several occasions in previous sections we will use the Wilson renormalization group recursion formula in order to study noncommutative phi-four at the self-dual point. As we will see in the next section a crucial step in this approach is truncation in which we replace every internal propagator $1/(p^2+\mu^2)$ by $1/(c^2+\mu^2)$ where $c$ is a constant and also replace every momentum loop integral by another constant. It will be clear in due course that these rules are essentially equivalent to the reduction of the effective action to zero dimension. In this section we will apply this reduction to the classical action    (\ref{model}) directly and study the resulting ordinary matrix model using the multi-trace approach of   \cite{O'Connor:2007ea}. The result for the $2-$point  proper vertex we will obtain in this section will provide a powerful check for our calculation using the Wilson recursion formula later.

 The reduction of the model (\ref{model}) to zero dimension is precisely the so-called Penner model (see below) which is a generalization of the hermitian quartic matrix model 
 \begin{eqnarray}
V_{0}(M)=Tr_N\big(rM^2+uM^4\big).
\end{eqnarray}
This is studied originally in the seminal work \cite{Brezin:1977sv}. The physics of this model is characterized by the celebrated one-cut to two-cut transition \cite{Shimamune:1981qf} which is expected and is known to persist at the level of  noncommutative scalar phi-four theory in all dimensions. Indeed  we can convince ourselves from Monte Carlo simulations that the one-cut to two-cut transition is lifted at the level of the noncommutative theory to the transition from disorder to the non-uniform or stripe phase \cite{Martin:2004un}. 

Our primary reason  in this article for applying the renormalization group idea to noncommutative phi-four is to probe this transition by means of an analytic non-perturbative method. We believe that the one-cut to two-cut transition can be accessed in a satisfactory way with the scheme  put forward in this article which consists in formulating noncommutative phi-four model in the matrix basis and employing the approximate renormalization group recursion formula and $1/N$ expansion. All this is based on the equivalence conjectured for example in \cite{Gubser:2000cd}  between noncommutative field theories at $\theta=\infty$ and large $N$ matrix models which is assumed to hold beyond the critical point. For an alternative viewpoint we refer to \cite{Bietenholz:2004as} who used a different regularization of noncommutative field theory based on the twisted reduced model. In any case in the disorder phase where both the symmetry $M\longrightarrow -M$ and translational symmetry are respected large $\theta$ noncommutative field theories are certainly equivalent to large $N$ matrix models.

The phase structure of  noncommutative phi-four is far more complicated \cite{Martin:2004un, Ambjorn:2002nj}. Besides the one-cut to two-cut transition (which becomes as we said the transition from disorder to non-uniform order) there is the usual Ising transition  (which is the transition from disorder to uniform order) and also a transition from a non-uniform order to a uniform order which all meet at a triple point.  As opposed to the usual uniform ordered phase where the symmetry $M\longrightarrow -M$ is spontaneously broken and translational symmetry is respected in the non-uniform ordered (stripe, two-cut) phase the symmetry $M\longrightarrow -M$ is maintained whereas translational symmetry is spontaneously broken. 

In non-commutative phi-four theory we have then three possible phases. The phase  characterized by the expectation value $<{M}>=0$, the phase characterized by  $<{M}>=\pm \sqrt{-r/2u}~{\bf 1}_N$  and the phase characterized by $<{M}>=\pm\sqrt{-r/2u}~\gamma$ where the idempotent $\gamma$ is given by $ \gamma=({\bf 1}_{N/2},-{\bf 1}_{N/2})$\footnote{There are other configurations in the non-uniform ordered phase proportional to other idempotents. We are here mentioning the configuration with the least action for brevity.}.  We use the terminology
\begin{eqnarray}
&&<{M}>=0~~{\rm disordered}~{\rm phase}.
\end{eqnarray}
\begin{eqnarray}
&&<{M}>=\pm\sqrt{-\frac{r}{2u}}{\bf 1}_N~~{\rm Ising}~({\rm uniform})~{\rm phase}.\label{ising}
\end{eqnarray}
\begin{eqnarray} 
&&<{M}>=\pm\sqrt{-\frac{r}{2u}}\gamma~~{\rm matrix}~({\rm non}~{\rm uniform}~{\rm or}~{\rm stripe})~{\rm phase}.\label{matrix}
\end{eqnarray}
There are therefore three possible phase transitions and as a consequence there must exist a triple point which has been confirmed in Monte Carlo simulation (see the second reference of \cite{Martin:2004un}). We have, from one hand,  the infamous $2$nd order Ising phase transition $0\longrightarrow \pm\sqrt{-r/2u}~{\bf 1}_N$ and the infamous $3$rd order matrix phase transition $0\longrightarrow \pm\sqrt{-r/2u}({\bf 1}_{N/2},-{\bf 1}_{N/2})$\footnote{The order of this transition at the level of non-commutative phi-four is still not clear. At the level of the matrix model it is $3$rd order.} while from the other hand there is also the  transition between the Ising and matrix configurations, viz ${\bf 1}_N\longrightarrow \gamma$, which was observed to be a continuation of the Ising line to large values of the coupling constant $u$ and thus it is also a $2$nd order phase transition.

In this article we certainly do not claim that we can access the three critical lines at once or equivalently the triple point at which they meet. This is undoubtedly  a very hard problem. We only claim that we can describe the  disordered-to-non-uniform-ordered (also called matrix) transition using the formalism put forward in this article. We remark that on both sides of this critical line the trace part of the matrix $M$ is close to zero (more on this in section $6$). 

The formalism of this article can not describe the Ising universality class fixed point since we are discussing physics in $d$ dimensions in terms of a critical behavior in $D=d-2$ dimensions and also it can not describe  the uniform-to-non-uniform-ordered transition because this transition does not respect the symmetry $M\longrightarrow -M$.  This symmetry was used in a crucial way in deriving the cumulant expansions  (\ref{mainresult}) and (\ref{primary}). The Ising model fixed point can still be accessed using the recursion formula along the lines  of \cite{Ydri:2012nw} while the  uniform-to-non-uniform-ordered transition may still be described by the current formalism if we do not insist on the symmetry $M\longrightarrow -M$. 
 
For an alternative analytical treatment of the three critical lines and the triple point at which they meet we refer the reader to the new article \cite{Polychronakos:2013nca}.

\subsection{The Multi-Trace Approach}
 The reduction of the model (\ref{model}) to zero dimension yields the pure matrix model (with $v_D$ the volume of momentum space)
\begin{eqnarray}
V&=&v_D Tr_N\bigg[\frac{1}{2}\mu^2M^+M+\frac{1}{2}r^2E\{M,M^+\}+\frac{v_D^2 u}{N}(M^+M)^2\bigg].
\end{eqnarray}
We will be interested in the real hermitian matrix model with $M^+=M$. By an appropriate rescaling we rewrite this as
\begin{eqnarray}
V_m&=&Tr_N\bigg[\frac{1}{2}M^2+m^2EM^2+\frac{g}{N}M^4\bigg].\label{penner}
\end{eqnarray}
The coupling constants $m^2$ and $g$ are given by
\begin{eqnarray}
m^2=\frac{r^2}{\mu^2}~,~g=\frac{v_Du}{(\mu^2)^2}.
\end{eqnarray}
 The path integral is given by
\begin{eqnarray}
Z_{m}=\int  d^{N^2}M~ \exp(-V_{m}(M)).
\end{eqnarray}
A very useful Schwinger-Dyson identity for this pure matrix model can be easily derived from the path integral
 \begin{eqnarray}
\frac{1}{Z_{m}}\int d^{N^2}M~\sum_{i,j=1}^{N}\frac{d}{dM_{ij}} \bigg(M_{ij}e^{-V_{m}(M)}\bigg)=0.
\end{eqnarray}
We obtain (with $G_2^m=<Tr_N M^2>_m/N^2$,   $G_e^m=<Tr_N EM^2>_m/N^3$ and $G_4^m=<Tr_N M^4>_m/N^3$)
\begin{eqnarray}
1=G_2^m+2m^2NG_e^m +4g G_4^m.\label{SD0}
\end{eqnarray}
We first consider the case $r^2=0$ or equivalently $m^2=0$. In this limit the model reduces to the famous hermitian quartic matrix model.
The partition function and the potential are given in this case by
\begin{eqnarray}
Z_{0}=\int  d^{N^2}M~ \exp(-V_{0}(M))~,~
V_{0}(M)=Tr_N\bigg(\frac{1}{2}M^2+\frac{g}{N}M^4\bigg).
\end{eqnarray}
The Schwinger-Dyson equation (\ref{SD0}) for $r^2=0$ can be put into the form (with $G_2^0=G_2$ and $G_4^0=G_4$)
\begin{eqnarray}
\Gamma_2=1+8gG_2-4g\Gamma_4(G_2)^3.\label{SD1}
\end{eqnarray}
The connected $2-$point and $4-$point functions and the $2-$point and  $4-$point proper vertices for this model are given by
\begin{eqnarray}
C_2=G_2~,~\Gamma_2=(C_2)^{-1}~,~G_2=\frac{1}{3}a^2(4-a^2)=1-8g_0+144g_0^2-3456g_0^3+....
\end{eqnarray}
\begin{eqnarray}
C_4=G_4-2(G_2)^2~,~\Gamma_4=-C_4(C_2)^{-4}~,~G_4=a^4(3-a^2).
\end{eqnarray}
Next we consider the case of small $r^2$. The partition function is given in this case by
\begin{eqnarray}
Z_{m}=\int  d^{N^2}M~ \exp(-V_{0}(M)-m^2Tr_N EM^2).
\end{eqnarray}
Now we will employ the method elaborated in \cite{O'Connor:2007ea} to calculate approximately this partition function.  Omitting the somewhat lengthy detail 
 we obtain an effective multi trace matrix model (correct up to order $m^2$) given by

\begin{eqnarray}
Z_{m}
&=&\int  d^{N^2}M~ \exp\big(-\frac{1}{2}(1+2Nm^2)Tr_N M^2-\frac{g}{N}Tr_NM^4+\frac{m^2}{2}(Tr_NM)^2\big).
\end{eqnarray}
We compute immediately
\begin{eqnarray}
<Tr_N EM^2>_m&=&-\partial_{m^2}\ln Z_{m}\nonumber\\
&=&N<Tr_N M^2>_{m}-\frac{1}{2}<(Tr_NM)^2>_{m}.
\end{eqnarray}
After a straightforward tedious calculation (see next section for a similar  calculation) we arrive at the $2-$point function (with $\hat{g}=g/(1+m^2N)^2$, $h=1+m^2N/2+...$)
\begin{eqnarray}
<M^2_{ij}>=\frac{{\delta}_{ij}N}{1+m^2N/2}\bigg[G_2(\hat{g},h)-\frac{m^2}{1+m^2N/2}(i-1/2)(1+h\partial_{h})G_2(\hat{g},h)+O(m^4)\bigg].\nonumber\\
\end{eqnarray}
\begin{eqnarray}
G_2(\hat{g},h)=1-8\hat{g}h+\hat{g}^2(64h+80h^2)+....
\end{eqnarray}
From this result we conclude that
\begin{eqnarray}
G_2^m=\frac{1}{1+m^2N/2}\bigg[G_2(\hat{g},h)+...\bigg].
\end{eqnarray}
\begin{eqnarray}
G_e^m=\frac{1}{2}\frac{1}{1+m^2N/2}\bigg[G_2(\hat{g},h)+...\bigg].
\end{eqnarray}
Thus at the leading order in $m^2$ we have $G_e^m=G_2^m/2$ and as a consequence the Schwinger-Dyson identity becomes $1=(1+m^2N)G_2^m+4gG_4^m$. This also means in particular that at the leading order in $m^2$ we have $<Tr_NJ^2>_m=0+O(m^2)$ where $J=M-Tr_NM/N$. The partition function can then be reduced further to
\begin{eqnarray}
Z_{m}
&=&\int  d^{N^2}M~ \exp\big(-\frac{1}{2}(1+Nm^2)Tr_N M^2-\frac{g}{N}Tr_NM^4-\frac{m^2N}{2}Tr_NJ^2\big)\nonumber\\
&=&\int  d^{N^2}M~ \exp\big(-\frac{1}{2}(1+Nm^2)Tr_N M^2-\frac{g}{N}Tr_NM^4\big).\label{Z}
\end{eqnarray}
Furthermore after a field scaling $M\longrightarrow M^{'}=(1+Nm^2/2)M$ we obtain $G_2^m=(1-Nm^2)G_2^{'}(\hat{g},Nm^2)$ and $G_4^m=(1-2Nm^2)G_4^{'}(\hat{g},Nm^2)$.  The primed correlation functions are computed with the partition function
\begin{eqnarray}
Z_{m}
&=&\int  d^{N^2}M^{'}~ \exp\big(-\frac{1}{2}Tr_N M^{'2}-\frac{\hat{g}}{N}Tr_NM^{'4}-\frac{m^2N}{2}Tr_NJ^{'2}\big).
\end{eqnarray}
 The Schwinger-Dyson identity becomes $1=G_2^{'}(\hat{g},Nm^2)+4\hat{g}G_4^{'}(\hat{g},Nm^2)$. In this form this identity holds true in an obvious way for $m^2=0$ since $G_2^{'}(\hat{g},0)=G_2^{}(g)$ and $G_4^{'}(\hat{g},0)=G_4^{}(g)$. We define the proper $2-$point vertex $\Gamma_2^{'}(\hat{g},Nm^2)$ by $\Gamma_2^{'}(\hat{g},Nm^2)=G_2^{'-1}(\hat{g},Nm^2)$. A straightforward calculation gives 
\begin{eqnarray}
\Gamma_2^{'}(\hat{g},Nm^2)&=&1+4\hat{g}G_4^{'}(\hat{g},Nm^2)G_2^{'-1}(\hat{g},Nm^2)\nonumber\\
&=&\Gamma_2^{'}(\hat{g},0)+4\hat{g}Nm^2\partial_{Nm^2}(G_4^{'}G_2^{'-1})|_{Nm^2=0}+...\nonumber\\
&=&\Gamma_2^{'}(\hat{g},0)+Nm^2\big(2(\Gamma_2^{'}-1)+\partial_{Nm^2}\Gamma_2^{'}|_{Nm^2=0}\big)+....
\end{eqnarray}
We rewrite this as
\begin{eqnarray}
(1+Nm^2)(\Gamma_2^{'}(\hat{g},Nm^2)-1)
&=&(1+Nm^2)(\Gamma_2^{'}(\hat{g},0)-1)+Nm^2\bigg[\Gamma_2^{'}-1\nonumber\\
&+&\partial_{Nm^2}\bigg((1+Nm^2)(\Gamma_2^{'}-1)\bigg)|_{Nm^2=0}\bigg].
\end{eqnarray}
Let us recall that $\Gamma_2^{'}(\hat{g},Nm^2)-1$ is the quantum correction to the mass of the field $M^{'}$ which is equal classically to exactly $1$. Now we undo the rescaling  $M\longrightarrow M^{'}=(1+Nm^2/2)M$. The quantum correction to the mass of the field  $M$ is $\Gamma_2(g,Nm^2)-1=(1+Nm^2)(\Gamma_2^{'}(\hat{g},Nm^2)-1)$ while the classical mass is $1+Nm^2$ as is obvious from (\ref{Z}). We obtain then
\begin{eqnarray}
\Gamma_2(g,Nm^2)
&=&\Gamma_2(g,0)+Nm^2\bigg[\Gamma_2-1+\partial_{Nm^2}\Gamma_2|_{Nm^2=0}\bigg]\nonumber\\
&=&\Gamma_2(g,0)+Nm^2\bigg[\Gamma_2-1+\frac{1}{2}\partial_{h}\Gamma_2|_{h=1}\bigg].
\end{eqnarray}
In the following section we will give, as a by product of our analysis of the noncommutative phi-four theory with $\omega=0$, a direct derivation of this formula together with an explicit perturbative expansion for $\partial_{Nm^2}\Gamma_2(g,Nm^2)$. In fact the above formulas will be used to test our method. We will also see in the following that the combination $\Gamma_e(g,0)=(1-\partial_{Nm^2}\Gamma_2(g,m^2))|_{m^2=0}$ is precisely the proper vertex associated with the operator $Tr_NEM^2$.

For a systematic more sophisticated study of the Penner matrix model (\ref{penner}) we refer the reader to the recent article \cite{Grosse:2012uv}.

\section{Wilson Renormalization Group }
\subsection{Recursion Formula}
In this subsection we review very briefly the original Wilson renormalization group recursion formula. By using the principle of truncation (more on this below) the action at any renormalization group step is taken to be given by 
\begin{eqnarray}
S_n(\phi)=\frac{1}{2}\int d^dx (\partial_{\mu}\phi_n)^2+\int d^dx P_n(\phi_n).\label{RF}
\end{eqnarray}
First we divide momenta logarithmically as  $1/2^l\leq |k|/\Lambda\leq 1/2^{l-1}$ where $\Lambda$ is the cutoff. Next by integrating out field modes with momenta in the highest shell  corresponding to $l=0$ and neglecting momentum dependence within each cell in position space (among other things) we arrive at an action of the same form with the replacements $\phi_n\longrightarrow \phi_{n+1}$ and $P_n\longrightarrow P_{n+1}$ where 
\begin{eqnarray}
Q_{n+1}(2^{d/2 }\alpha_n^{-1}z)&\equiv &w^{-1}P_{n+1}(\phi_{n+1})=-2^d\ln \frac{I_n(z)}{I_n(0)}.
\end{eqnarray} 
The variable $z$ is related to the field $\phi_{n+1}$ in a particular way which does not interest us in this article whereas $w^{-1}$ is the volume of a single cell in position space and $\alpha_n$ is the normalization of the field. Indeed the fields $\phi_n$ and $\phi_{n+1}$ are related by 
\begin{eqnarray}
\phi_n(x)=\Phi_n(x)+2^{-d/2}\alpha_n\phi_{n+1}(x/2). 
\end{eqnarray} 
The background field $\Phi_n$ contains precisely the integrated momenta  $1\leq |k|/\Lambda\leq 2$. The function $I_n(z)$ appearing in the recursion formula (\ref{RF}) is given by the integral
\begin{eqnarray}
I_n(z)=\int dy~\exp\bigg(-y^2-\frac{1}{2}Q_n(y_{}+z_{})-\frac{1}{2}Q_n(-y_{}+z_{})\bigg).\label{i0}
\end{eqnarray}
\subsection{Cumulant Expansion}
The Wilson renormalization group idea consists in general in the following three main steps:
\begin{itemize}
\item{}{\bf Integration and Truncation:} Define a new (renormalized) action $S^{'}$ by integrating out all modes with momenta larger or equal than $\rho\Lambda$ where $0\leq \rho\leq 1$ and $\Lambda$ is the cutoff. This is a very complicated step which requires in practice the use of truncation and many other approximations before it can be carried out explicitly. Wilson truncation means that we calculate quantum corrections to only those terms which appear in the starting action. The principal approximation we will further use consists in the above Wilson recursion formula (\ref{RF}). We state without proof (see \cite{Wilson:1973jj} for a derivation) that the use of this recursion formula is completely equivalent to the use in perturbation theory of the Polyakov-Wilson rules   given by the following two approximations:

\begin{itemize}
\item{}We replace every internal propagator  $1/(k^2+\mu^2)$ by $1/(c^2+\mu^2)$ where $c$ is a constant usually taken to be $\Lambda$.
\item{} We replace every momentum loop integral $\int_{\rho\Lambda}^{\Lambda} d^Dk/(2\pi)^D$ by another constant $v_D=\Lambda^D\hat{v}_D$ where 
\begin{eqnarray}
\hat{v}_D=\frac{2(1-\rho^D)}{D}\frac{1}{(4\pi)^{{D}/{2}}}\frac{1}{\Gamma({D}/{2})}.
\end{eqnarray}
\end{itemize}   
We remark here that the above two rules are rather obvious and in fact natural in the limit  of the dilatation parameter given by $\rho\longrightarrow 1$ regardless of the Wilson recursion formula itself.
\item{}{\bf Rescaling:} This a simple step which consists in restoring the cutoff to the original value $\Lambda$ which can be achieved by simply making the change of variables $p\longrightarrow p^{'}=\rho p$.  
\item{}{\bf Normalization:} Restoring the standard normalization of the kinetic term which is equal to $1/2$ by appropriately rescaling the field variable. This is also a very simple step.  
\end{itemize}
Let us then start by decomposing  the $N\times N$ matrix $M$ into an $N\times N$ background matrix $\tilde{M}$ and an $N\times N$ fluctuation matrix $m$, viz
\begin{eqnarray}
M=\tilde{M}+m.
\end{eqnarray}
The background $\tilde{M}$ contains slow modes, i.e. modes with momenta less or equal than $\rho\Lambda$ while the fluctuation $m$  contains fast modes, i.e. modes with momenta  larger  than $\rho\Lambda$ where $0<\rho<1$. By integrating out the field $m$ we obtain the partition function

\begin{eqnarray}
Z
&=&\int d\tilde {M}e^{-S[\tilde{M}]}~e^{-\Delta S(\tilde{M})}.
\end{eqnarray}
By using momentum conservation, the symmetry $m\longrightarrow -m$ and also by neglecting non-planar diagrams (in the sense of the $1/N$ expansion) we find that the non-perturbative correction $\Delta S(\tilde{M})$   is given explicitly (up to the fourth power in the field $\tilde{M}$) by the cumulant expansion  
\begin{eqnarray}
\Delta S(\tilde{M}) &=&4\frac{u}{N}\int d^Dx <Tr_N\tilde{M}^2m^2(x)>_{\rm co}\nonumber\\
&-&8\frac{u^2}{N^2}\int d^Dx\int d^Dy <Tr_N\tilde{M}m^3(x)Tr_N\tilde{M}m^3(y)>_{\rm co}\nonumber\\
&-&8\frac{u^2}{N^2}\int d^Dx\int d^Dy <Tr_N\tilde{M}^2m^2(x)Tr_N\tilde{M}^2m^2(y)>_{\rm co}\nonumber\\
&+&32 \frac{u^3}{N^3}\int d^Dx\int d^Dy \int d^D z <Tr_N\tilde{M}m^3(x)Tr_N\tilde{M}m^3(y)Tr_N\tilde{M}^2m^2(z)>_{\rm co}\nonumber\\
&-&\frac{32}{3} \frac{u^4}{N^4}\int d^Dx\int d^D y\int d^Dz\int d^Dw<Tr_N\tilde{M}m^3(x)Tr_N\tilde{M}m^3(y)Tr_N\tilde{M}m^3(z)Tr_N\tilde{M}m^3(w)>_{\rm co}.\label{mainresult}\nonumber\\
\end{eqnarray}
A derivation of this fundamental result is given in the second appendix. The notation "${\rm co}$" stands for the connected component. The first and second terms yield correction to the mass parameter whereas the last three terms yield correction to the quartic coupling constant. The wave function renormalization is obtained from the expansion around $p^2=0$ of the second term. This is the most difficult contribution to calculate  as we will see shortly.
\subsection{Mass and Harmonic Oscillator Renormalizations}
Quantum corrections to the mass parameter $\mu^2$ and to the harmonic oscillator coupling constant $r^2$ are obtained from the first term of (\ref{mainresult}) and also from the second term  of (\ref{mainresult}) evaluated at $p^2=0$, viz
\begin{eqnarray}
\Delta S_{\rm mass+h.o}&=&\frac{4 u}{N}\int d^Dx <Tr_N\tilde{M}^2m^2(x)>_{\rm co}-\frac{8u^2}{N^2}\int d^Dx\int d^Dy <Tr_N\tilde{M}m^3(x)Tr_N\tilde{M}m^3(y)>_{\rm co}|_{p^2=0}.\nonumber\\
\end{eqnarray}
 The corresponding Feynman diagrams are shown on figures (\ref{figure1}) and (\ref{figure2}). We included diagrams up to three loops. The series can be continued if one wishes to go further. However all the subleading (in $\lambda$) Feynman diagrams which dominate the large $N$ limit can be recovered in our scheme as we will now show. By using first the approximations (\ref{appr1}) and (\ref{appr2}) and then the Wilson recursion formula as described above the first set of Feynman diagrams shown on figure (\ref{figure1}) reduces to

\begin{eqnarray}
4\frac{u}{N}\int d^Dx <Tr_N\tilde{M}^2m^2(x)>_{\rm co}&=&\int d^Dx \tilde{M}^2(x)_{ii}\bigg[4 v_D u {\Delta}_{n_0i}(c)-32 (v_D u)^2 {\Delta}_{n_0i}^2(c){\Delta}_{n_0n_0}(c)\nonumber\\
&+ & (v_D u)^3\bigg(256{\Delta}_{n_0i}^2(c){\Delta}^3_{n_0n_0}(c)+320{\Delta}_{n_0i}^3(c){\Delta}_{n_0n_0}^2(c)\bigg)+....\bigg].\label{massR1}\nonumber\\
\end{eqnarray}
By expanding the second set of Feynman diagrams shown on figure (\ref{figure2}) around $p^2=0$ and retaining the $0$ th order term  we get extra corrections to the mass and harmonic oscillator. By using again the approximations (\ref{appr1}) and (\ref{appr2}) and the Wilson recursion formula we get explicitly 

\begin{eqnarray}
&&-8\frac{u^2}{N^2}\int d^Dx\int d^Dy <Tr_N\tilde{M}m^3(x)Tr_N\tilde{M}m^3(y)>_{\rm co}|_{p^2=0}=\int d^Dx \tilde{M}(x)_{ij}\tilde{M}(x)_{ji}\nonumber\\
&\times &\bigg[-8(v_D u)^2{\Delta}_{n_0i}(c){\Delta}_{n_0j}(c){\Delta}_{n_0n_0}(c)+256 (v_D u)^3{\Delta}_{n_0i}(c){\Delta}_{n_0j}^2(c){\Delta}_{n_0n_0}^2(c)+...\bigg].\label{massR2}\nonumber\\
\end{eqnarray}
The terms multiplying  $\tilde{M}^2(x)_{ii}$ and $\tilde{M}(x)_{ij}\tilde{M}(x)_{ji}$  still depend on the indices $i$ and $j$ which is the source of the harmonic oscillator renormalization. Recall that the harmonic oscillator term is of the form $\int d^Dx \tilde{M}^2(x)_{ii}(i-1/2)$. The mass correction corresponds to setting these indices equal to some fixed value $n$ whereas the harmonic oscillator correction corresponds to expanding these terms linearly around $n$. Again in the spirit of Wilson contraction we treat the index $i$ (or $j$) as a continuous  variable and expand the propagator around $i=n$ as  follows

\begin{eqnarray}
{\Delta}_{in_0}(c)&=&{\Delta}_{nn_0}(c)-(i-n)r^2{\Delta}_{nn_0}^2(c)+(i-n)^2r^4{\Delta}_{nn_0}^3(c)+...\label{pro}
\end{eqnarray}
This actually makes sense since we are assuming that $\theta$ is sufficiently large and thus $r^2$ is sufficiently small. 

The full mass and harmonic oscillator renormalizations are contained in the sum of the two terms (\ref{massR1}) and (\ref{massR2}). By using the expansion (\ref{pro}) we obtain after some calculation the result

\begin{eqnarray}
\Delta S_{\rm mass+h.o}
&=&\int d^Dx\bigg[\frac{1}{2}(c^2+\mu^2+r^2N)\Gamma_2(g,h)  Tr_N\tilde{M}^2(x)+\frac{1}{2}r^2\partial_{\kappa}\Gamma_2(g,h) Tr_N(E-n+\frac{1}{2})\tilde{M}^2(x)\nonumber\\
&-&\frac{n}{2}\frac{r^4}{c^2+\mu^2}\partial_{\kappa}^2\Gamma_2(g,h)Tr_N E\tilde{M}^2(x)
\bigg]+...
\end{eqnarray}
This formula contains the leading as well as the subleading corrections in $1/\theta$ to the mass and to the harmonic oscillator terms. The $2-$point proper vertex $\Gamma_2(g,h)$ is defined by
\begin{eqnarray}
\Gamma_2(g,h)=8gh-80g^2h^2+2g^3 (256h^2+576h^3)+....
\end{eqnarray}
The new coupling constants $g$ and $h\equiv 1/\kappa$ are defined by
\begin{eqnarray}
g=v_D u {\Delta}_{n_0n_0}^2(c)~,~h=\frac{{\Delta}_{nn_0}(c)}{{\Delta}_{n_0n_0}(c)}.
\end{eqnarray}
At $\theta=\infty$ we have $h=1$ and thus $\Gamma_2(g,1)=8g-80g^2+1664g^3+....$. The $2-$point proper vertex $\Gamma_2(g,1)$ is expected to be related to the $2-$point proper vertex  $\Gamma_2(g)$ of the quartic matrix model, i.e. the expansion $8g-80g^2+1664g^3+....$ should be recognized from the pure quartic matrix model. Indeed we have
\begin{eqnarray}
\Gamma_2(g)-1=\Gamma_2(g,1).
\end{eqnarray}
The $2-$point proper vertex  $\Gamma_2(g)$ is in fact a function known non perturbatively   given by (with $a=(\sqrt{1+48g}-1)/{24g}$)
\begin{eqnarray}
\Gamma_2(g)=\frac{3}{a^2(4-a^2)}=1+8g-80g^2+1664g^3-....
\end{eqnarray}
The full  mass and harmonic oscillator action is therefore given by (using also $h=1+r^2(n_0-n)/(c^2+\mu^2)+...$)

\begin{eqnarray}
S_{\rm mass+h.o}+\Delta S_{\rm mass+h.o}
&=&\frac{1}{2}\bigg\{(\mu^2+c^2)\Gamma_2(g)-c^2+r^2 N\big(\Gamma_2(g)-\Gamma_e(g)\big)+O(r^4)\bigg\} \int d^Dx Tr_N\tilde{M}^2(x)\nonumber\\
&+ &  r^2  \bigg\{\Gamma_e(g)- \frac{r^2(N+1)}{2(c^2+\mu^2)}\partial_{\kappa}\Gamma_e(g,h)|_{\kappa=1}+O(r^4)\bigg\} \int d^DxTr_NE\tilde{M}^2(x).\label{r1}\nonumber\\
\end{eqnarray}
We have defined
\begin{eqnarray}
\Gamma_e(g,h)&=&1+\frac{1}{2}\partial_{\kappa}\Gamma_2(g,h)~,~\Gamma_e(g)=\Gamma_e(g,1).
\end{eqnarray}
In contrast to $\Gamma_2(g)$ the function $\Gamma_e(g)$ is only known perturbatively. Note also that we did not need to fix the value of $n$ in deriving the result (\ref{r1}). In the following we will see more clearly the virtue of the choice $n=1/2$ which was assumed in a previous section on pure matrix models. For $c^2=0$ the mass renormalization agrees with the result obtained from the Penner matrix model.

\subsection{Wave Function Renormalization}
The wave function renormalization is also obtained from the  $2$nd term of (\ref{mainresult}) and as a consequence the relevant Feynman diagrams are still given by those shown on figure (\ref{figure2}). More precisely we need, as before, to expand these diagrams around $p^2=0$ but retain now the linear term in $p^2$ which is very difficult to do explicitly. We will present here a new estimation of the coefficient of $p^2$ which is consistent with the recursion formula.

The three graphs shown explicitly on figure (\ref{figure2}) are of the form
\begin{eqnarray}
\int \frac{d^Dp_1}{(2\pi)^D}\int \frac{d^Dp_2}{(2\pi)^D}\Delta_{in_0}(p+p_1+p_2)f(p_1,p_2).\label{int}
\end{eqnarray}
We need the expansion around $p^2=0$ of the propagator ${\Delta}_{in_0}(p+p_1+p_2)=1/((p_1+p_2+p)^2+\sigma^2)$ where $\sigma^2=\mu^2+r^2(i+n_0-1)$. By using rotational invariance we can replace $\Delta_{in_0}(p+p_1+p_2)$ in this integral as follows (with $\epsilon=4-D$ and $\epsilon_0=\epsilon/D$)

\begin{eqnarray}
{\Delta}_{in_0}(p+p_1+p_2)={\Delta}_{in_0}(p_1+p_2)+\big[\epsilon_0(p_1+p_2)^2-\sigma^2\big]{\Delta}_{in_0}^3(p_1+p_2)p^2+....\label{ar1}
\end{eqnarray}
The first term corresponds to the $p^2=0$ contribution relevant for mass and harmonic oscillator renormalizations whereas the second term will provide wave function renormalization.  The contribution of this last term to the integral (\ref{int}) is given by
\begin{eqnarray}
p^2\int \frac{d^Dp_1}{(2\pi)^D}\int \frac{d^Dp_2}{(2\pi)^D}\big[\epsilon_0(p_1+p_2)^2-\sigma^2\big]{\Delta}_{in_0}^3(p_1+p_2)f(p_1,p_2).
\end{eqnarray}
In the recursion formula we replace the internal momentum appearing in each propagator by an average value $c$ which we have chosen to be $c=\Lambda$. 
We anticipate the application of the recursion formula and replace the factor ${\epsilon}_0(p_1+p_2)^2 - \sigma^2$ with $\epsilon_0c^2-\sigma^2$. We get then the estimation
\begin{eqnarray}
p^2\big[\epsilon_0 c^2-\sigma^2\big]\int \frac{d^Dp_1}{(2\pi)^D}\int \frac{d^Dp_2}{(2\pi)^D}{\Delta}_{in_0}^3(p_1+p_2)f(p_1,p_2).\label{appr3}
\end{eqnarray}
In higher order diagrams there could be more than one propagator depending on $p$ and also the propagators may depend on more than just  two internal momenta which will complicate the situation even further. More precisely  if we generalize (\ref{ar1}) to higher order terms than the extra factor ${\epsilon}_0(p_1+p_2)^2 - \sigma^2$ may be different. This important issue will be discussed further below.

The graphs shown on figure (\ref{figure2}) then yields the result (using also the approximations (\ref{appr1}) and (\ref{appr2}) and the recursion formula)
\begin{eqnarray}
&-&8\frac{u^2}{N^2}\int d^Dx\int d^Dy <Tr_N\tilde{M}m^3(x)Tr_N\tilde{M}m^3(y)>_{\rm co}=
\int \frac{d^Dp}{(2\pi)^D}p^2 \bar{M}_{ij}(p)\bar{M}_{ji}(-p)\nonumber\\
&\times &\bigg[\mu^2+r^2(i+n_0-1)-\epsilon_0 c^2\big]\bigg]\bigg\{8(v_Du)^2{\Delta}_{n_0i}^3(c){\Delta}_{n_0j}(c){\Delta}_{n_0n_0}(c)-192 (v_Du)^3\nonumber\\
&\times & {\Delta}_{n_0i}^3(c){\Delta}_{n_0n_0}^2(c){\Delta}_{n_0j}^2(c)-64 (v_Du)^3{\Delta}_{n_0i}^3(c){\Delta}_{n_0j}^2(c){\Delta}_{n_0n_0}^2(c)+ ....\bigg\}.
\end{eqnarray}
We split the factor $p^2[\mu^2+r^2(i+n_0-1)-\epsilon_0c^2]$ as $p^2[\mu^2+r^2(n_0-1/2)-\epsilon_0c^2]$ plus $p^2r^2(i-1/2)$.  The first bit  depends on the indices $i$ and $j$ only through the propagators. Now we only need the contribution obtained by setting these indices equal to the fixed value $n=1/2$. The linear term around $n=1/2$ will lead to a correction of the form $Tr_N E (\partial_{\mu} M)^2$ which is not present in the original action. For $n\neq 1/2$ the correction will be of the form $Tr_N (E+{\rm constant}) (\partial_{\mu} M)^2$. By the same reasoning we discard the correction  $\bar{M}_{ij}(p)\bar{M}_{ji}(-p)(p^2r^2(i-1/2))$ since it corresponds to a correction of the form $Tr_N E (\partial_{\mu} M)^2$. This is the principle of Wilson truncation in action again. We get then the result
\begin{eqnarray}
&-&8\frac{u^2}{N^2}\int d^Dx\int d^Dy <Tr_N\tilde{M}m^3(x)Tr_N\tilde{M}m^3(y)>_{\rm co}=
\int \frac{d^Dp}{(2\pi)^D}p^2 \bar{M}_{ij}(p)\bar{M}_{ji}(-p)\nonumber\\
&\times &[\mu^2+r^2(n_0-\frac{1}{2})-\epsilon_0c^2]\bigg\{8(v_Du)^2{\Delta}_{n_0n}^4(c){\Delta}_{n_0n_0}(c)-256 (v_Du)^3{\Delta}_{n_0n}^5(c){\Delta}_{n_0n_0}^2(c)\nonumber\\
&+& ....\bigg\}.
\end{eqnarray}
We rewrite this as
\begin{eqnarray}
&-&8\frac{u^2}{N^2}\int d^Dx\int d^Dy <Tr_N\tilde{M}m^3(x)Tr_N\tilde{M}m^3(y)>_{\rm co}=\frac{\mu^2+{r^2N}/{2}-\epsilon_0c^2}{\mu^2+{r^2N}/{2}+c^2}\big[Z_2(g,1)\nonumber\\
&+&r^2\frac{N}{2(c^2+\mu^2)} \partial_hZ_2(g,h)|_{h=1}+O(r^4)\big]\int d^Dx Tr_N(\partial_{\mu}\tilde{M})^2.
\end{eqnarray}
We have defined
\begin{eqnarray}
Z_2(g,h)=8g^2h^3-256g^3h^4+....
\end{eqnarray}
We need now to find a combination of Green's functions and proper vertices of the pure matrix model with an expansion given exactly by $8g^2-256g^3+...$. From the Schwinger-Dyson equation (\ref{SD1}) we propose that the function $2g\Gamma_4(g)G_2^3(g)$ is the correct guess. Notice the resemblance of  the graphs corresponding to $8gG_2$ and $-4g\Gamma_4G_2^3$ to the graphs associated with the terms  $4(u/N)\int d^Dx <Tr\tilde{M}^2m^2(x)>_{\rm co}$ and $-8(u^2/N^2)\int d^Dx\int d^Dy <Tr\tilde{M}m^3(x)Tr\tilde{M}m^3(y)>_{\rm co}$ respectively. Indeed we compute
\begin{eqnarray}
Z_2(g)\equiv Z_2(g,1)&=&2g\Gamma_4(g)G_2^3(g)\nonumber\\
&=&\frac{2g}{3}\frac{a^2(1-a^2)(5-2a^2)}{4-a^2}\nonumber\\
&=&8g^2-256g^3+...
\end{eqnarray}
The full kinetic action is given by
\begin{eqnarray}
 S_{\rm kinetic}+\Delta S_{\rm kinetic}
&=&\bigg\{\frac{1}{2}+\frac{{\mu}^{2}-\epsilon_0 c^2}{c^2+\mu^2}Z_2(g)+\frac{r^2N}{2(c^2+\mu^2)^2}\big(c^2(1+\epsilon_0)Z_2(g)+({\mu}^{2}-\epsilon_0 c^2)\nonumber\\
&\times &\partial_hZ_2(g,h)|_{h=1}\big)+O(r^4)\bigg\}\int d^DxTr_N(\partial_{\mu}\tilde{M})^2.\label{r2}
\end{eqnarray}
As mentioned previously in higher order diagrams there could be more than one propagator depending on $p$ and also these propagators may depend on more than just  two internal momenta and as a consequence the extra factor ${\epsilon}_0(p_1+p_2)^2 - \sigma^2$ appearing in  (\ref{appr3}) may be different. This coefficient leads to the crucial factor $({{\mu}^{2}-\epsilon_0 c^2})/({c^2+\mu^2})$ which  multiplies  the non perturbative function $Z_2(g)$. The central conjecture of this section is that within the scheme of the recursion formula outlined above the coefficient $({{\mu}^{2}-\epsilon_0 c^2})/({c^2+\mu^2})$ is in fact the same coefficient appearing in higher order corrections of the function $Z_2(g,h)$.

In order to introduce a non trivial wave function renormalization we can alternatively employ the simpler approximation motivated by dimensional consideration used in the context of matrix models in \cite{Ferretti:1995zn,Nishigaki:1996ts}.  In this scheme the first derivative of the propagator with respect to the external momentum $p^2$ will be approximated by the multiplication with the given propagator as follows
 \begin{eqnarray}
p^2\frac{d}{dp^2}(...)|_{p^2=0}&=&-p^2\frac{1}{(p_1+p_2+p)^2+\sigma^2}(...)|_{p^2=0}.\label{ferretti}
\end{eqnarray}
In this case the integral (\ref{int}) is approximation with
\begin{eqnarray}
-p^2\int \frac{d^Dp_1}{(2\pi)^D}\int \frac{d^Dp_2}{(2\pi)^D}\Delta_{in_0}^2(p_1+p_2)f(p_1,p_2).\label{appr4}
\end{eqnarray}
The difference between (\ref{appr3}) and (\ref{appr4}) consists in the difference between the factors 
 \begin{eqnarray}
\big[\sigma^2-\epsilon_0\Lambda^2\big]{\Delta}_{in_0}(p_1+p_2)|_{\ref{appr3}}\leftrightarrow 1|_{\ref{appr4}}.
\end{eqnarray}
Finally we also mention here  Golner extension of the recursion formula derived in  \cite{Golner:1973zz} which also allows a non trivial wave function renormalization. As it turns out this is a more consistent framework which requires however a non trivial modification of the starting action and thus we do not discuss it any further here.

\subsection{Quartic Interaction Renormalization}
This is obtained from the last three terms of  equation (\ref{mainresult}) and without any further calculation we simply state the final result obtained using the approximation (\ref{appr1}) and (\ref{appr2}) and the recursion formula.  This is given by
\begin{eqnarray}
\Delta S_{\rm interaction}
&=&\frac{u}{N}[\frac{{\Gamma}_4(g,h)}{4g}-1]\int d^Dx Tr_N\tilde{M}^4.
\end{eqnarray}
The function $\Gamma_4(g,h)$ is now defined by the expansion
\begin{eqnarray}
\frac{{\Gamma}_4(g,h)}{4g}=1-8gh^2+g^2(160h^4+64h^3)+....
\end{eqnarray}
The full interaction term is therefore
\begin{eqnarray}
S_{\rm interaction}+\Delta S_{\rm interaction}
&=&\frac{u}{N}\frac{{\Gamma}_4(g,h)}{4g}\int d^Dx Tr_N\tilde{M}^4(x)\nonumber\\
&=&\frac{u}{N}\frac{1}{4g}\bigg\{{\Gamma}_4(g,1)+\frac{r^2N}{2(c^2+\mu^2)}\partial_h{\Gamma}_4(g,h)|_{h=1}+O(r^4)\bigg\}\int d^Dx Tr_N\tilde{M}^4(x).\label{r3}\nonumber\\
\end{eqnarray}
As expected the $4-$point proper vertex ${\Gamma}_4(g,1)$ is precisely the $4-$point proper vertex of the quartic matrix model given by the expansion 
\begin{eqnarray}
\Gamma_4(g,1)\equiv \Gamma_4(g)=\frac{9(1-a^2)(5-2a^2)}{a^4(1-a^2)^4}=4g-32g^2+896g^3+...
\end{eqnarray}

\subsection{Renormalization Group Equations}
So far we have only gone through the first step of the Wilson renormalization group approach, namely integration and truncation. By putting the results (\ref{r1}), (\ref{r2}) and (\ref{r3}) we obtain the effective action including all leading and subleading corrections in $1/\theta$ at the leading order in $1/N$. By performing the second step of the Wilson renormalization group approach, i.e. by scaling momenta as $p\longrightarrow p/\rho$ so that the cutoff returns to its original value $\Lambda$ we obtain the effective action


\begin{eqnarray}
 S+\Delta S
&=&\frac{\rho^{2+D}}{2}\bigg\{ Z(g,\mu^2)+\frac{r^2N}{(c^2+\mu^2)^2}\Delta Z(g,\mu^2) +O(r^4)\bigg\}\int_0^{\Lambda} \frac{d^Dp}{(2\pi)^D} p^2 Tr_N\bar{M}^*(\rho p)\bar{M}(\rho p)\nonumber\\
&+&\frac{\rho^D}{2}\bigg\{\Gamma(g,\mu^2)+r^2 N\Delta \Gamma(g)+O(r^4)\bigg\} \int_0^{\Lambda} \frac{d^Dp}{(2\pi)^D}  Tr_N\bar{M}^*(\rho p)\bar{M}(\rho p)\nonumber\\
&+ &  r^2 \rho^D \bigg\{\Gamma_e(g)+ \frac{r^2(N+1)}{c^2+\mu^2}\Delta \Gamma_e(g)+O(r^4)\bigg\} \int_0^{\Lambda} \frac{d^Dp}{(2\pi)^D}  Tr_NE\bar{M}^*(\rho p)\bar{M}(\rho p)\nonumber\\
&+&\frac{u\rho^{3D}}{N}\frac{1}{4g}\bigg\{{\Gamma}_4(g)+\frac{r^2N}{c^2+\mu^2}\Delta {\Gamma}_4(g)+O(r^4)\bigg\} \int_0^{\Lambda} \frac{d^Dp_1}{(2\pi)^D}...  \int_0^{\Lambda} \frac{d^Dp_4}{(2\pi)^D} Tr_N\bar{M}(\rho p_1)...\bar{M}(\rho p_4)\nonumber\\
&\times & (2\pi)^D\delta^D(p_1+...p_4).\label{rescaling}
\end{eqnarray}
The functions $Z(g,\mu^2)$, $\Gamma(g,\mu^2)$, $\Gamma_2(g)$ and $\Gamma_4(g)$ are known non perturbatively whereas we were able to determine the functions $\Delta Z(g,\mu^2)$, $\Delta \Gamma(g)$, $\Delta \Gamma_4(g)$, $\Delta \Gamma_e(g)$ and $\Gamma_e(g)$ only perturbatively.  These functions are summarized in table (\ref{table}). 

The third and final step of the Wilson renormalization group approach consists in rescaling the field in such a way that  the kinetic term is brought to its canonical form as follows
\begin{eqnarray}
\bar{M}^{'}(p)=\rho^{\frac{2+D}{2}}\sqrt{Z(g,\mu^2)+\frac{r^2N}{(c^2+\mu^2)^2}\Delta Z(g,\mu^2)}\bar{M}(\rho p).
\end{eqnarray}
We get the action to the form
\begin{eqnarray}
 S+\Delta S&=&\frac{1}{2}\int_0^{\Lambda} \frac{d^Dp}{(2\pi)^D} p^2 Tr_N\bar{M}^{'*}(p)\bar{M}^{'}(p)\nonumber\\
&+&\frac{1}{2}\frac{\rho^{-2}}{Z(g,\mu^2)}\bigg[\Gamma(g,\mu^2)+r^2 N\Delta \Gamma(g)-\frac{r^2N}{(c^2+\mu^2)^2}\frac{\Gamma(g,\mu^2)\Delta Z(g,\mu^2)}{Z(g,\mu^2)}+O(r^4)\bigg]\nonumber\\
&\times &  \int_0^{\Lambda} \frac{d^Dp}{(2\pi)^D}  Tr_N\bar{M}^{'*}(p)\bar{M}^{'}(p)\nonumber\\
&+&\frac{\rho^{-2}r^2}{Z(g,\mu^2)}\bigg[\Gamma_e(g)+\frac{r^2(N+1)}{c^2+\mu^2}\Delta \Gamma_e(g)-\frac{r^2N}{(c^2+\mu^2)^2}\frac{\Gamma_e(g)\Delta Z(g,\mu^2)}{Z(g,\mu^2)}+O(r^4)\bigg]\nonumber\\
&\times &\int_0^{\Lambda} \frac{d^Dp}{(2\pi)^D}  Tr_NE\bar{M}^{'*}(p)\bar{M}^{'}(p)\nonumber\\
&+&\frac{u}{N}\frac{1}{4g}\frac{\rho^{-\epsilon}}{Z^2(g,\mu^2)}\bigg[{\Gamma}_4(g)+\frac{r^2N}{c^2+\mu^2}\Delta\Gamma_4(g)-\frac{2r^2N}{(c^2+\mu^2)^2}\frac{\Gamma_4(g)\Delta Z(g,\mu^2)}{Z(g,\mu^2)}+O(r^4)\bigg]  \nonumber\\
&\times &  \int_0^{\Lambda} \frac{d^Dp_1}{(2\pi)^D}...  \int_0^{\Lambda} \frac{d^Dp_4}{(2\pi)^D}Tr_N\bar{M}^{'}(p_1)...\bar{M}^{'}(p_4)(2\pi)^D\delta^D(p_1+...p_4).\label{effectiveactionSD2}
\end{eqnarray}
Let us recall that $r^2$ is related to the harmonic oscillator coupling constant $\Omega$ by $\nu_2r^2=4\pi(\Omega^2+1)$ and that we have started with an action at the self-dual point, i.e. $\Omega^2=1$ or equivalently $\omega=0$. Thus the renormalization of $r^2$ can be understood either as a renormalization of $\Omega^2$ or as a renormalization of the noncommutativity parameter. Both descriptions must be equivalent but we prefer to think that  we have always $\Omega^2=1$ and attribute all the renormalization to $\theta$.

In position space the above action reads as follows 
\begin{eqnarray}
 S+\Delta S&=&\frac{1}{2}\int d^Dx Tr_N(\partial_{\mu}\tilde{M}^{'})^2+\frac{\mu^{'2}}{2}\int d^Dx Tr_N\tilde{M}^{'2}+r^{'2}\int d^Dx Tr_NE \tilde{M}^{'2}\nonumber\\
&+&\frac{u^{'}}{N}\int d^Dx Tr_N\tilde{M}^{'4}.
\end{eqnarray}
The renormalized mass ${\mu}^{'2}$, the renormalized quartic coupling constant ${u}^{'}$ and the renormalized  inverse noncommutativity $r^{'2}$ are given by 
\begin{eqnarray}
{\mu}^{'2}=\frac{\rho^{-2}}{Z(g,\mu^2)}\bigg[\Gamma(g,\mu^2)+r^2 N\Delta \Gamma(g)-\frac{r^2N}{(c^2+\mu^2)^2}\frac{\Gamma(g,\mu^2)\Delta Z(g,\mu^2)}{Z(g,\mu^2)}+O(r^4)\bigg].
\end{eqnarray}
\begin{eqnarray}
r^{'2}=\frac{\rho^{-2}r^2}{Z(g,\mu^2)}\bigg[\Gamma_e(g)+\frac{r^2(N+1)}{c^2+\mu^2}\Delta \Gamma_e(g)-\frac{r^2N}{(c^2+\mu^2)^2}\frac{\Gamma_e(g)\Delta Z(g,\mu^2)}{Z(g,\mu^2)}+O(r^4)\bigg].
\end{eqnarray}
\begin{eqnarray}
 {u}^{'}=u\frac{\rho^{-\epsilon}}{Z^2(g,\mu^2)}\frac{1}{4g}\bigg[{\Gamma}_4(g)+\frac{r^2N}{c^2+\mu^2}\Delta\Gamma_4(g)-\frac{2r^2N}{(c^2+\mu^2)^2}\frac{\Gamma_4(g)\Delta Z(g,\mu^2)}{Z(g,\mu^2)}+O(r^4)\bigg]. 
\end{eqnarray}
The process which led from the bare coupling constants  $\mu^2$, $r^2$ and $u$ to the renormalized coupling constants ${\mu}^{'2}$, $r^{'2}$ and $u^{'}$ can be repeated an arbitrary number of times. The bare coupling constants will be denoted by $\mu_{0}^{2}$, $r_{0}^{2}$ and $u_{0}$ whereas the the renormalized coupling constants  at the first step of the renormalization group procedure will be denoted by  $\mu_{1}^{2}$, $r_{1}^{2}$ and $u_{1}$. At a generic step $l+1$ of the renormalization group process the renormalized coupling constants  $\mu_{l+1}^{2}$, $r_{l+1}^{2}$ and $u_{l+1}$ are related to their previous values $\mu_{l}^{2}$, $r_{l}^{2}$ and $u_{l}$ by the renormalization group equations 

\begin{eqnarray}
{\mu}_{l+1}^{2}=\frac{\rho^{-2}}{Z(g_l,\mu_l^2)}\bigg[\Gamma(g_l,\mu_l^2)+r_l^2 N\Delta \Gamma(g_l)-\frac{r_l^2N}{(c^2+\mu_l^2)^2}\frac{\Gamma(g_l,\mu_l^2)\Delta Z(g_l,\mu_l^2)}{Z(g_l,\mu_l^2)}\bigg].\label{rgf1}
\end{eqnarray}
\begin{eqnarray}
r_{l+1}^{2}=\frac{\rho^{-2}r_l^2}{Z(g_l,\mu_l^2)}\bigg[\Gamma_e(g_l)+\frac{r_l^2(N+1)}{c^2+\mu_l^2}\Delta \Gamma_e(g_l)-\frac{r_l^2N}{(c^2+\mu_l^2)^2}\frac{\Gamma_e(g_l)\Delta Z(g_l,\mu_l^2)}{Z(g_l,\mu_l^2)}\bigg].\label{rgf2}
\end{eqnarray}
\begin{eqnarray}
 {u}_{l+1}^{}=u_l\frac{\rho^{-\epsilon}}{Z^2(g_l,\mu_l^2)}\frac{1}{4g}\bigg[{\Gamma}_4(g_l)+\frac{r_l^2N}{c^2+\mu_l^2}\Delta\Gamma_4(g_l)-\frac{2r_l^2N}{(c^2+\mu_l^2)^2}\frac{\Gamma_4(g_l)\Delta Z(g_l,\mu_l^2)}{Z(g_l,\mu_l^2)}\bigg].\label{rgf3} 
\end{eqnarray}
The effective coupling constant $g_l$ will of course be given in terms of  $\mu_{l}^{2}$, $r_{l}^{2}$ and $u_{l}$ by the same formula that related $g_0$ to $\mu_{0}^{2}$, $r_{0}^{2}$ and $u_{0}$. We are therefore interested in renormalization group flow in a $3-$dimensional parameter space generated by the mass $\mu^2$, the quartic coupling constant $u$ and the harmonic oscillator coupling constant (inverse noncommutativity) $r^2$.

\section{Fixed Points and Critical Exponents}
\subsection{The Matrix Model Fixed Point} 
 By definition a renormalization group fixed point is a point in the space parameter which is invariant under the renormalization group flow. If we denote the fixed point by $\mu_*^2$, $r_{*}^2$ and $u_*$     then we must have \footnote{It should be noted that looking for the fixed point $u_*$ of the coupling constant $u$ is equivalent to looking for zeros of the beta function.}
\begin{eqnarray}
{\mu}_{*}^{2}=\frac{\rho^{-2}}{Z(g_*,\mu_*^2)}\bigg[\Gamma(g_*,\mu_*^2)+r_*^2 N\Delta \Gamma(g_*)-\frac{r_*^2N}{(c^2+\mu_*^2)^2}\frac{\Gamma(g_*,\mu_*^2)\Delta Z(g_*,\mu_*^2)}{Z(g_*,\mu_*^2)}\bigg].\label{mu*}
\end{eqnarray}
\begin{eqnarray}
r_{*}^{2}=\frac{\rho^{-2}r_*^2}{Z(g_*,\mu_*^2)}\bigg[\Gamma_e(g_*)+\frac{r_*^2(N+1)}{c^2+\mu_*^2}\Delta \Gamma_e(g_*)-\frac{r_*^2N}{(c^2+\mu_*^2)^2}\frac{\Gamma_e(g_*)\Delta Z(g_*,\mu_*^2)}{Z(g_*,\mu_*^2)}\bigg].\label{r*}
\end{eqnarray}
\begin{eqnarray}
 {u}_{*}^{}=u_*\frac{\rho^{-\epsilon}}{Z^2(g_*,\mu_*^2)}\frac{1}{4g_*}\bigg[{\Gamma}_4(g_*)+\frac{r_*^2N}{c^2+\mu_*^2}\Delta\Gamma_4(g_*)-\frac{2r_*^2N}{(c^2+\mu_*^2)^2}\frac{\Gamma_4(g_*)\Delta Z(g_*,\mu_*^2)}{Z(g_*,\mu_*^2)}\bigg].\label{u*} 
\end{eqnarray}
The second equation is new by comparison with the commutative theory. The definition of $g_*$ in terms of $\mu_*^2$, $r_{*}^2$ and $u_*$ is obvious. There are possibly several soultions (fixed points) of interest to these renormalization group equations. We will mainly concentrate on the matrix model fixed point corresponding to infinite noncommutativity which is the most obvious solution to equation (\ref{r*}) given by
\begin{eqnarray}
{r}_{*}^2=0.
\end{eqnarray}
The remaining two equations reduce then to
\begin{eqnarray}
{\mu}_{*}^{2}=\frac{\rho^{-2}}{Z(g_*,\mu_*^2)}\Gamma(g_*,\mu_*^2).\label{mu*1}
\end{eqnarray}

\begin{eqnarray}
 {u}_{*}^{}=u_*\frac{\rho^{-\epsilon}}{Z^2(g_*,\mu_*^2)}{\Gamma}_4(g_*).\label{u*1}
\end{eqnarray}
Thus this fixed point is fully determined by functions which are known non-perturbatively. An obvious solution to (\ref{u*1}) is $u_*=0$ which corresponds to the usual Gaussian fixed point. By discarding this solution equation (\ref{u*1}) becomes
 \begin{eqnarray}
 1=\frac{\rho^{-\epsilon}}{Z^2(g_*,\mu_*^2)}\frac{{\Gamma}_4(g_*)}{4g_*}.\label{u*2}
\end{eqnarray}
We define the new variable 
\begin{eqnarray}
 \alpha=\frac{\mu^2}{c^2+\mu^2}.
\end{eqnarray}
The solutions of equations (\ref{mu*1}) and (\ref{u*2}) are given respectively by
\begin{eqnarray}
&&\alpha_{*}=\frac{f_1(\rho,g_*)+2\epsilon_0Z_2(g_*)}{ 4(1+\epsilon_0)Z_2(g_{*})}.
\end{eqnarray}
\begin{eqnarray}
\alpha_{*}=\frac{f_2(\rho,g_*)+2\epsilon_0Z_2(g_*)}{ 4(1+\epsilon_0)Z_2(g_{*})}.
\end{eqnarray}
The functions $f_1$ and $f_2$ are defined by
\begin{eqnarray}
&& f_1=-1+\frac{1}{\rho^2}- \sqrt{(1-\frac{1}{\rho^2}-2\epsilon_0 Z_2(g_{*}))^2+\frac{8}{\rho^2}(1+\epsilon_0)Z_2(g_{*})(\Gamma_2(g_{*})-1)}.\nonumber\\
\end{eqnarray}
\begin{eqnarray}
&&f_2=-2- 2\sqrt{\rho^{-\epsilon}\frac{{\Gamma}_4(g_{*})}{4g_{*}}}+2Z_2(g_{*})\epsilon_0.
\end{eqnarray}
Clearly we must have $f_1=f_2$. In other words $g_*$ is the solution of the consistency equation
\begin{eqnarray}
f=f_1-f_2\equiv 0.\label{consistency}
\end{eqnarray}
The physical region of $g$ is $[0,\infty[$ while the full domain of definition is  $[-1/48,\infty[$. Furthermore the functions $\Gamma_2$, $\Gamma_4$  and $Z_2$ depend on $g$ only through  $a=a(g)$  defined by $a^2={2}/({\sqrt{1+48 g}+1})$.  Graphically we observe that the two functions $f_1$ and $f_2$ intersect for all dimensions $d=2,3,4$. We will employ the Newton-Raphson algorithm to determine the roots of the consistency condition (\ref{consistency}). The roots are then given by the iterative equation (with an initial step which in principle can be anything)
 \begin{eqnarray}
 x_{i+1}=x_i-\frac{f(x_i)}{df(x_i)/dx_i}.
\end{eqnarray}
After a sufficient number of iterations $n$ we will obtain the root $x_n\equiv g_{*}$. We determine $\alpha_*$, the critical mass $\hat{\mu}_*^2=\mu_*^2/\Lambda^2$ and the critical coupling $\hat{u}_*=u_*  \Lambda^{D-4}$ from the equations 

\begin{eqnarray}
 \alpha_{*}=\frac{f_2+2\epsilon_0 Z_2(g_{*})}{4(1+\epsilon_0)Z_2(g_{*})}.
\end{eqnarray}
 \begin{eqnarray}
\hat{\mu}_*^2=\frac{\alpha_*}{1-\alpha_*}.
\end{eqnarray}
\begin{eqnarray}
\hat{u}_*=\frac{g_{*}(1+\hat{\mu}_*^2)^2}{\hat{v}_D}.
\end{eqnarray}
The results of this calculation are shown on table (\ref{table1}). For comparison we report the critical values in the approximation (\ref{appr4})  in table (\ref{table2}).  There is of course in each dimension the extra Gaussian fixed point as we have discussed. There is only the Gaussian fixed point in $D=4(d=6)$ in the approximation (\ref{appr4}) which can not be true since the matrix model fixed point is expected to exist in all dimensions. In our approximation  (\ref{appr3}) we have checked that there is always a non-trivial fixed point for any value of $\rho$ in the interval $0<\rho<1$. 

In the remainder we will compute the mass critical exponent $\nu$ and the anomalous dimension $\eta$ within this scheme. 
\subsection{Theoretical Digression} 
The computation of the mass critical exponent $\nu$ requires linearization of the renormalization group equations (\ref{rgf1}), (\ref{rgf2}) and (\ref{rgf3}).  Let us first rewrite these renormalization group equations in the compact form
 \begin{eqnarray}
G^{(n+1)}={\cal M}(G^{(n)},\rho).
\end{eqnarray}
The vector of coupling constants $G$ is defined by $G=(G_1,G_2,G_3)$ where  $G_1=\mu^2$, $G_2=u$ and  $G_3=r^2$. The linearized renormalization group equations are of the form (with $\delta G=G-G_*$)
 \begin{eqnarray}
\delta G^{(n+1)}={M}(G_*,\rho)\delta G^{(n)}.
\end{eqnarray}
The matrices $M$ and ${\cal M}$ are different for different fixed points. The matrix $M$ is related to ${\cal M}$ as
 \begin{eqnarray}
M_{ij}(G_*,\rho)=\frac{\partial}{\partial G_j}{\cal M}_j(G^{(n)},\rho)|_{G^{(n)}=G_*}.
\end{eqnarray}
Let $v_{\alpha}^T(\rho)$ be the left eigenvectors of $M$ with eigenvalues $\lambda_{\alpha}(\rho)$, viz
  \begin{eqnarray}
v_{\alpha}^T(\rho)M(G_*,\rho)=\lambda_{\alpha}(\rho)M(G_*,\rho).
\end{eqnarray}
The so-called scaling fields $u_{\alpha}$ are defined by the projection of the vector of coupling constants $\delta G$ onto the left  eigenvectors $v_{\alpha}^T(\rho)$, viz
 \begin{eqnarray}
u_{\alpha}(\rho)=v_{\alpha}^T(\rho)\delta G.
\end{eqnarray}
These vectors do not mix under the renormalization group transformations since they satisfy 
\begin{eqnarray}
u_{\alpha}^{(n+1)}(\rho)=\lambda_{\alpha}(\rho)u_{\alpha}^{(n)}(\rho).
\end{eqnarray}
Thus the scaling field $u_{\alpha}$ will increase under renormalization group transformations (relevant) if $\lambda_{\alpha}>1$ while it will decrease if $\lambda_{\alpha}<1$ (irrelevant).

It turns out that every real asymmetric matrix $M$ can be factored (Schur factorization) as $M=U^{-1}TU$ where $T$ is upper triangular and $U$ is unitary. This substitutes for diagonalization which is not always possible for real asymmetric matrices. The eigenvalues in this case are real or come in complex conjugate pairs. As usual the eigenvalues are  the zeros of the characteristic polynomial. 

A fundamental property of the renormalization group is the semigroup property which for two different dilatation parameters $\rho_1$ and $\rho_2$ reads
\begin{eqnarray}
M(G_*,\rho_1)M(G_*,\rho_2)=M(G_*,\rho_2)M(G_*,\rho_1)=M(G_*,\rho_1\rho_2).
\end{eqnarray}
It is not difficult to convince ourselves that this leads to the requirement that the renormalization group eigenvalues $\lambda_{\alpha}(\rho)$ scale as
\begin{eqnarray}
\lambda_{\alpha}(\rho)=\lambda_{\alpha}(1)\rho^{-y_{\alpha}}.\label{beha}
\end{eqnarray}
The so-called scaling indices $y_{\alpha}$ (also called  critical exponents) are the  renormalization group eigenvalues of the scaling fields $u_{\alpha}(\rho)$. Indeed under a renormalization group transformation we have $u_{\alpha}(\rho)\longrightarrow u^{'}_{\alpha}(\rho)=\lambda_{\alpha}(\rho)u_{\alpha}(\rho)$ and thus for an infinitesimal renormalization group transformation ($\rho\longrightarrow 1$) we must have $u^{'}_{\alpha}(\rho)-\lambda_{\alpha}(1)u_{\alpha}(\rho)=-y_{\alpha}\ln \rho \lambda_{\alpha}(1)u_{\alpha}(\rho)+O((\ln \rho)^2)$. In other words

\begin{eqnarray}
\rho\frac{d}{d\rho}u_{\alpha}(\rho)=-y_{\alpha}u_{\alpha}(\rho).
\end{eqnarray}
In the limit $\rho\longrightarrow 0$ we have $u_{\alpha}(\rho)\longrightarrow \pm \infty$ when $y_{\alpha}>0$ (relevant) and  $u_{\alpha}(\rho)\longrightarrow \pm \infty$ when $y_{\alpha}<0$ (irrelevant).

\subsection{The Mass Critical Exponent $\nu$}
\begin{figure}[htbp]
\begin{center}
\includegraphics[width=5cm,angle=-90]{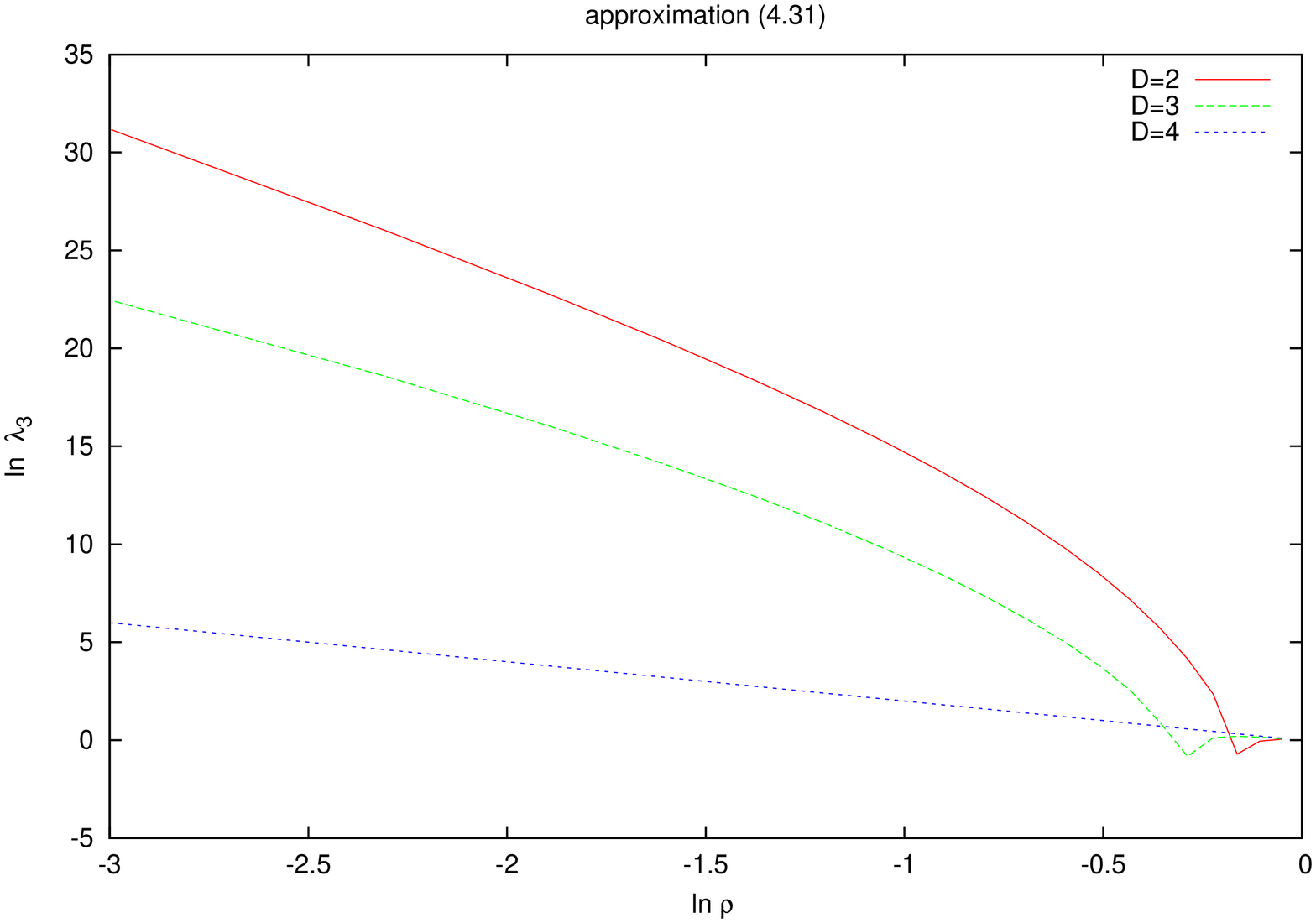}
\includegraphics[width=5cm,angle=-90]{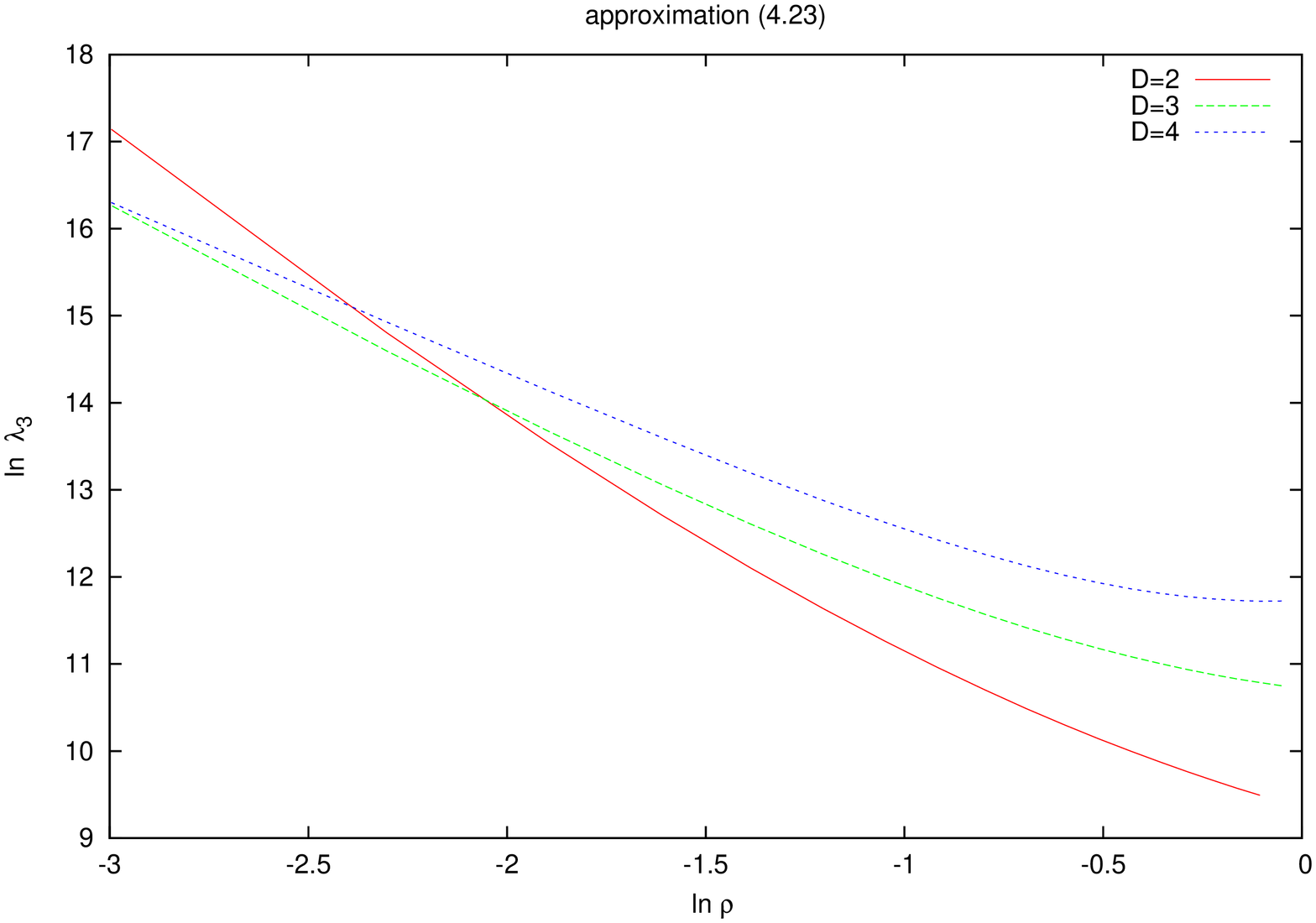}
\caption{The eigenvalues $\lambda_3$ in the approximations  (\ref{appr3}) and  (\ref{appr4}) respectively. 
}\label{graphs0}
\end{center}
\end{figure}

The matrix $M$ in our case is of the form
\begin{eqnarray}
\left( \begin{array}{ccc}
 M_{11} & M_{12} & M_{13}\\
 M_{21} & M_{22} & M_{23} \\
0 & 0 & M_{33}
\end{array} \right).\label{rgm}
\end{eqnarray}
The corresponding characteristic polynomial gives the two equations
\begin{eqnarray}
M_{33}-\lambda=0~,~{\rm det} \left( \begin{array}{cc}
 M_{11}-\lambda & M_{12} \\
 M_{21} & M_{22}-\lambda 
\end{array} \right)=0.
\end{eqnarray}
The eigenvalue in the direction $G_3=r^2$ is therefore given by
\begin{eqnarray}
{\lambda}_3=M_{33}=\frac{\rho^{-2}\Gamma_e(G_*)}{Z(G_*,\mu_*^2)}.
\end{eqnarray}
This eigenvalue (or more precisely $ \ln \lambda_3$)  is plotted on figure (\ref{graphs0}) as a function of $\ln \rho$. It is immediately obvious that $\lambda_3>1$ and hence $r^2$ is a relevant coupling constant like the mass. However the function $\Gamma_e(g)$ used in the above formula is only known perturbatively and hence this conclusion should be taken with care. 

The two remaining eigenvalues are determined from the linearized renormalization group equations in the $2-$dimensional space generated by $G_1=\mu^2$ and $G_2=u$. As it turns out this problem depends only on functions which are fully known non perturbatively. The eigenvalues $\lambda_1$ and $\lambda_2$ can be determined from the trace and determinant which are given by
  \begin{eqnarray}
\lambda_1+\lambda_2=M_{11}+M_{22}\equiv {\rm Tr}_2 M~,~\lambda_1\lambda_2=M_{11}M_{22}-M_{12}M_{21}\equiv {\rm det}_2M.
\end{eqnarray}
In other words
 \begin{eqnarray}
\lambda_1=\frac{{\rm Tr}_2 M\pm \sqrt{({\rm Tr}_2M)^2-4{\rm det}_2M}}{2}~,~\lambda_2={\rm Tr}_2 M-\lambda_1.
\end{eqnarray}
These are real as long as $({\rm Tr}_2M)^2-4{\rm det}_2M \geq 0$. Indeed a real asymmetric matrix can have in general complex eigenvalues. For a discussion of the relevance and interpretation of complex renormalization group  eigenvalues  see chapter $3$ of \cite{Kopietz:2010zz} and references therein. This point is not discussed in this article.

The formula (\ref{beha}) was used as a crucial test for our numerical calculations. In particular we have determined by means of this formula the range of the dilatation parameter $\rho$ over which the logarithm  of the eigenvalues scale  linearly with $\ln \rho$. It is natural to expect this behavior to hold only if the renormalization group steps are sufficiently small so not to alter drastically the infrared physics of the problem. 

\begin{figure}[htbp]
\begin{center}
\includegraphics[width=5cm,angle=-90]{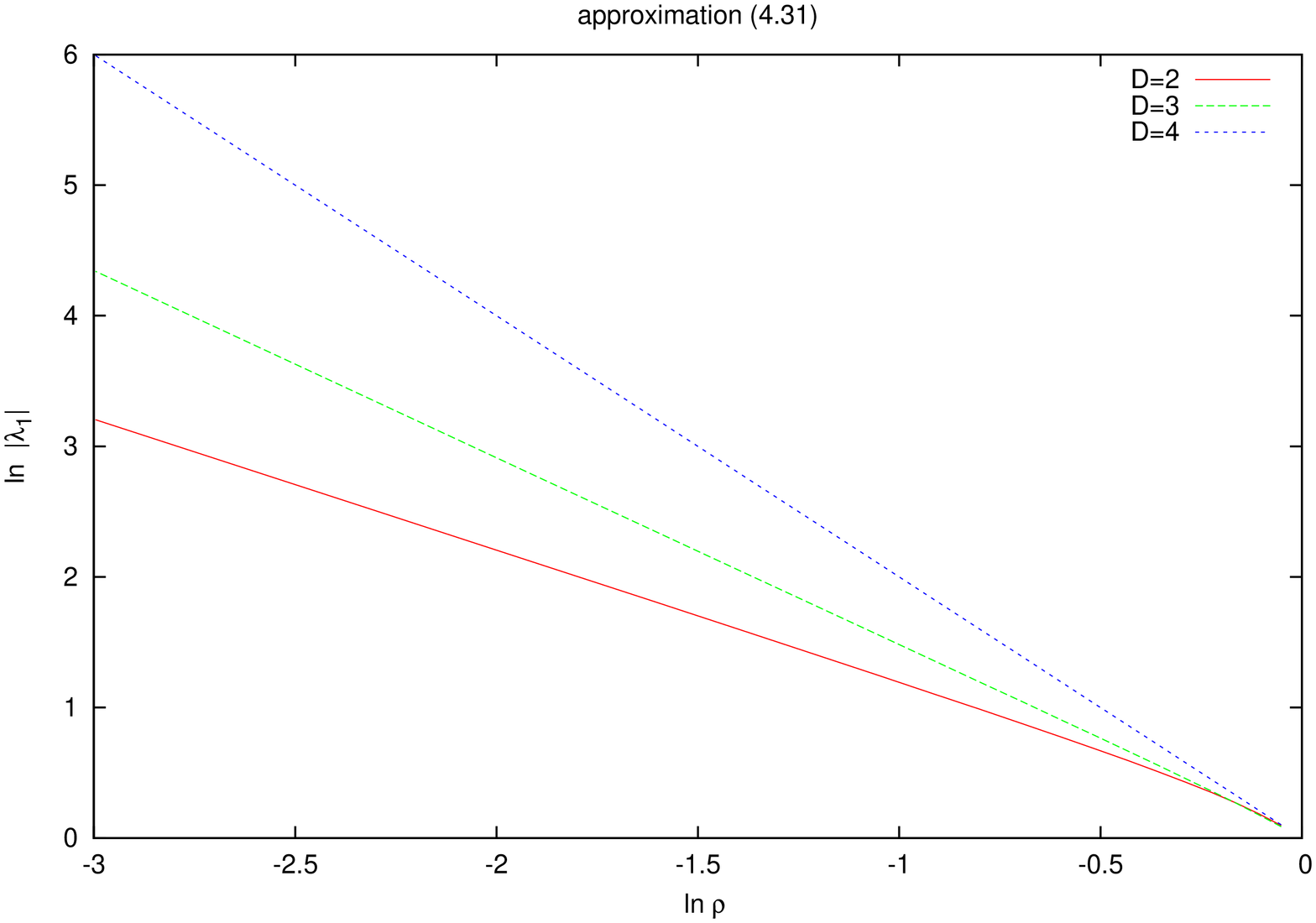}
\includegraphics[width=5cm,angle=-90]{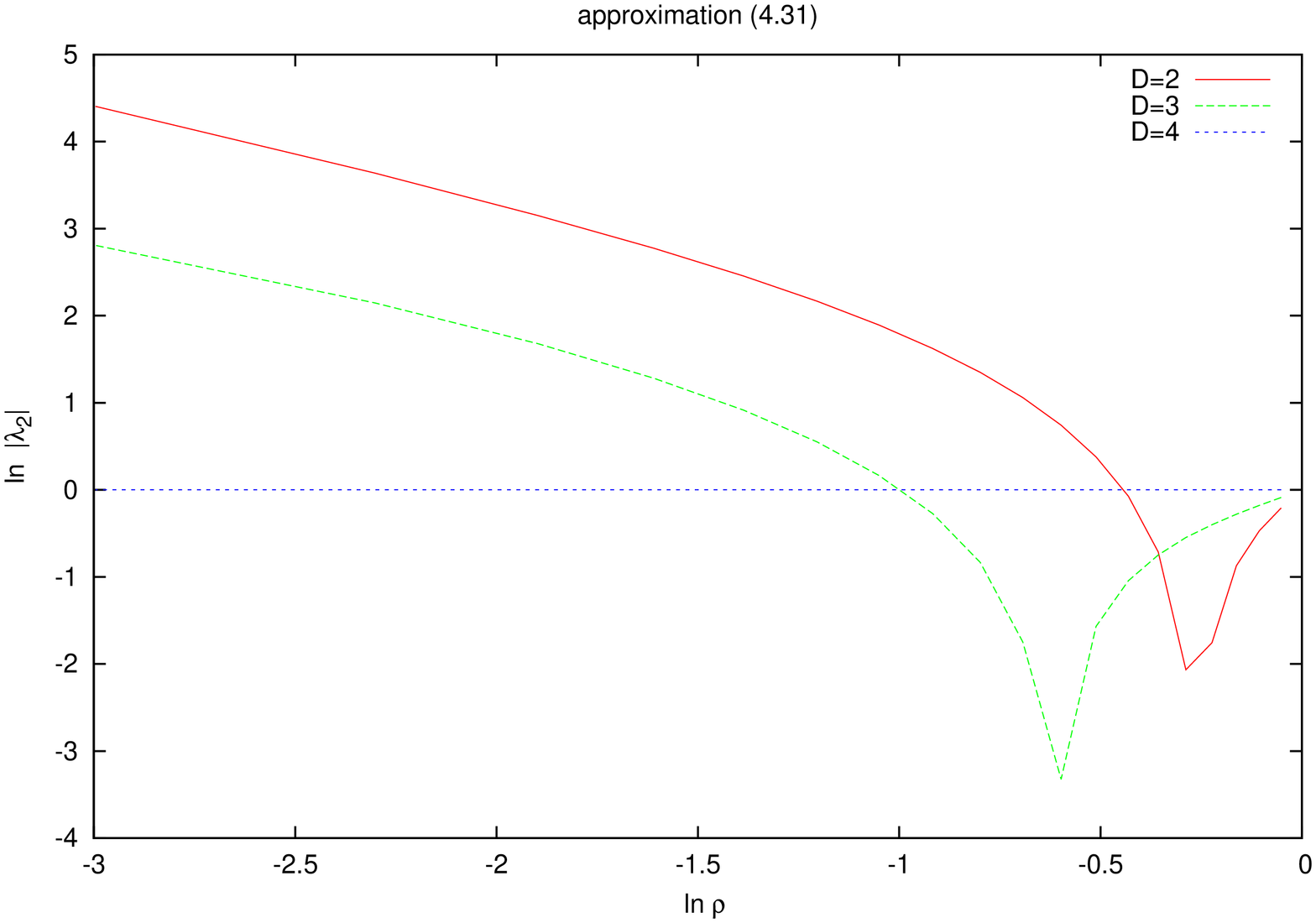}
\includegraphics[width=5cm,angle=-90]{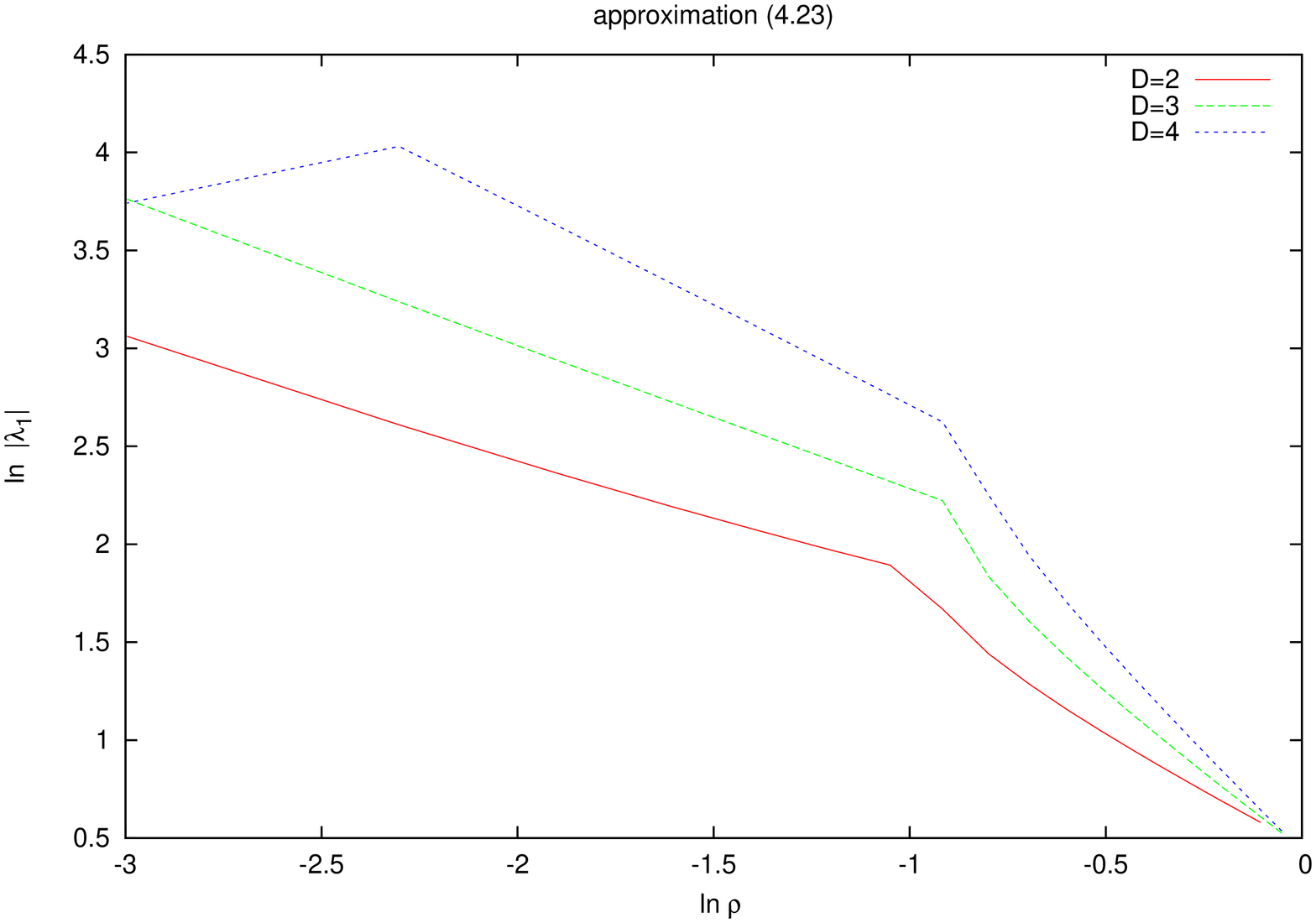}
\includegraphics[width=5cm,angle=-90]{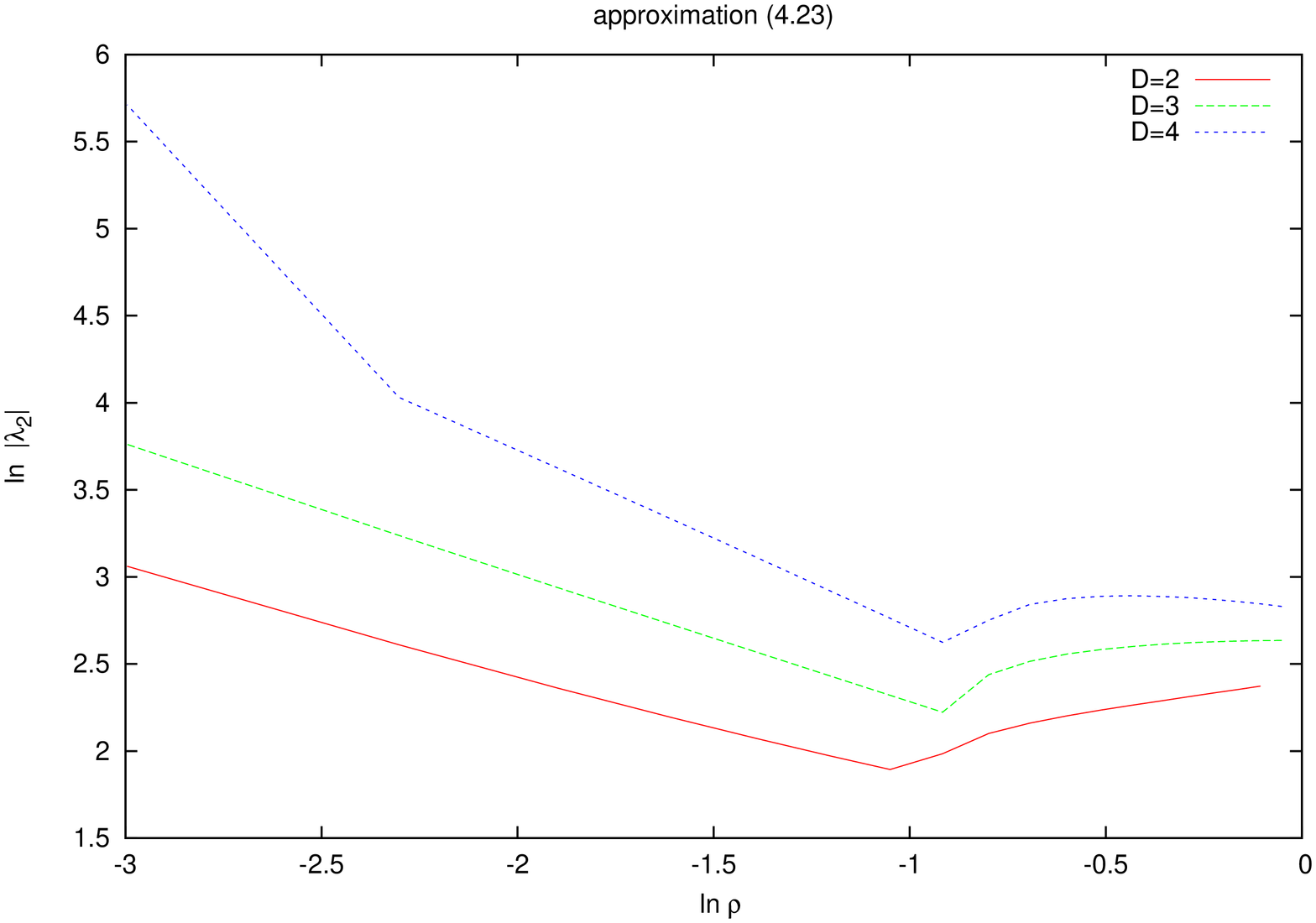}
\caption{The eigenvalues $\lambda_1$  and $\lambda_2$ in the approximations  (\ref{appr3}) and  (\ref{appr4}) respectively.  }\label{graphs}
\end{center}
\end{figure}

Some results of this test are shown on figure (\ref{graphs}). In the approximation  (\ref{appr3}) the eigenvalues $\lambda_1$ and $\lambda_2$  become complex and conjugate to each other at around $\ln \rho\simeq -1$ and thus the regime $\ln \rho >-1$ is of no interest to us here. In the regime $\ln \rho < -1$ both $\lambda_1$ and $\lambda_2$ are real. The eigenvalue   $\ln |\lambda_{1}|$ as a function of $\ln\rho$ is linear in the  regime $\ln \rho < -1$ whereas the eigenvalue   $\ln |\lambda_{2}|$ is linear in this regime only for $D=2$. For $D=3$ and $4$ $\ln |\lambda_{2}|$ is linear only for $\ln \rho << -1$ which is not very satisfactory. Fortunately it is the eigenvalue $\lambda_1$ which is associated with a relevant scaling field and thus used to define the mass critical exponent as we will discuss shortly.

Note that the eigenvalues $\lambda_1$ and $\lambda_2$ are both negative in the regime  $\ln \rho < -1$ which means that the scaling indices $y_1$ and $y_2$ are mildly complex. Furthermore we note that over the  regime $\ln \rho < -1$ the eigenvalue $\lambda_3$ scales also correctly with $\rho$ (figure (\ref{graphs})).

\begin{figure}[htbp]
\begin{center}
\includegraphics[width=5cm,angle=-90]{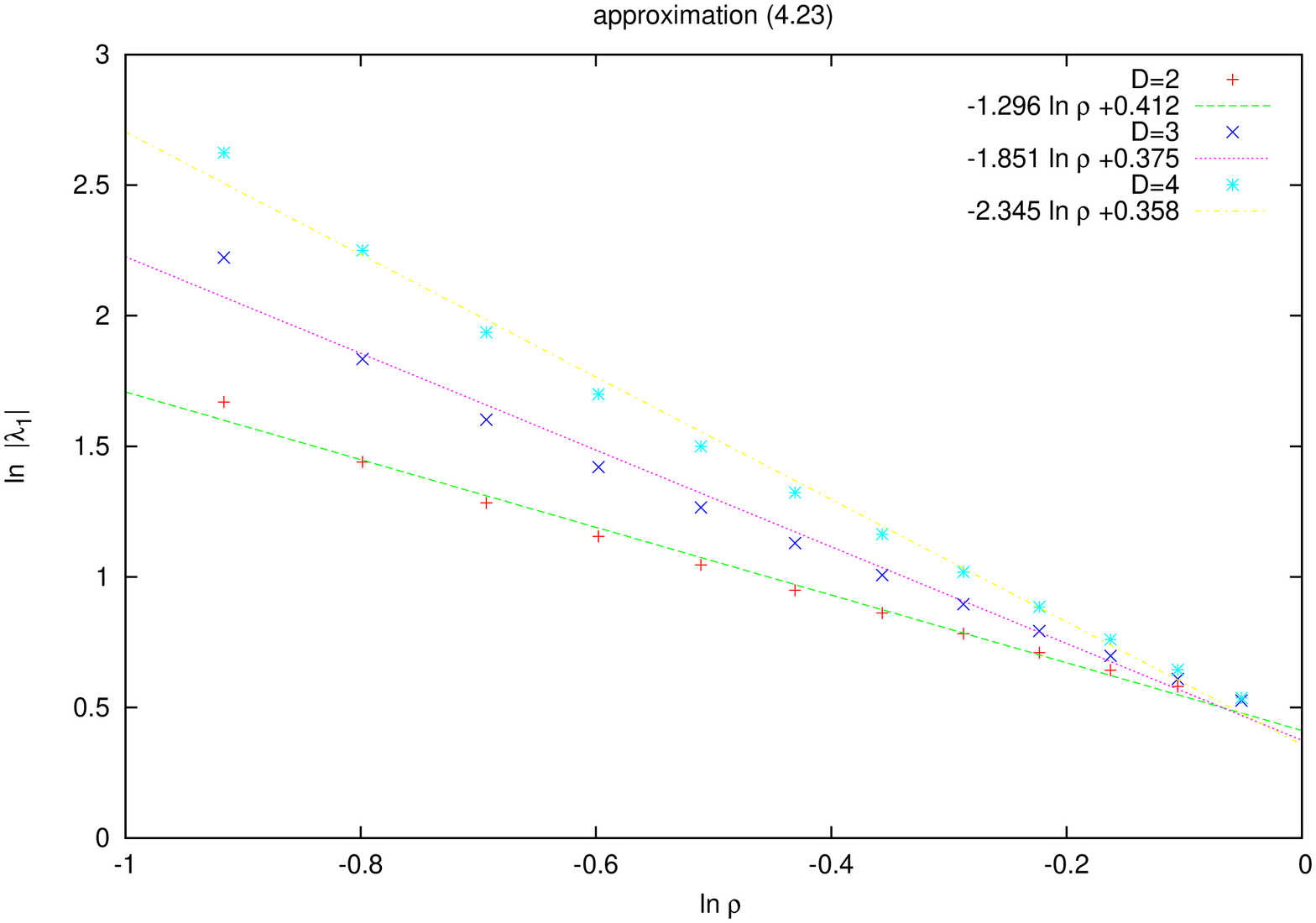}
\includegraphics[width=5cm,angle=-90]{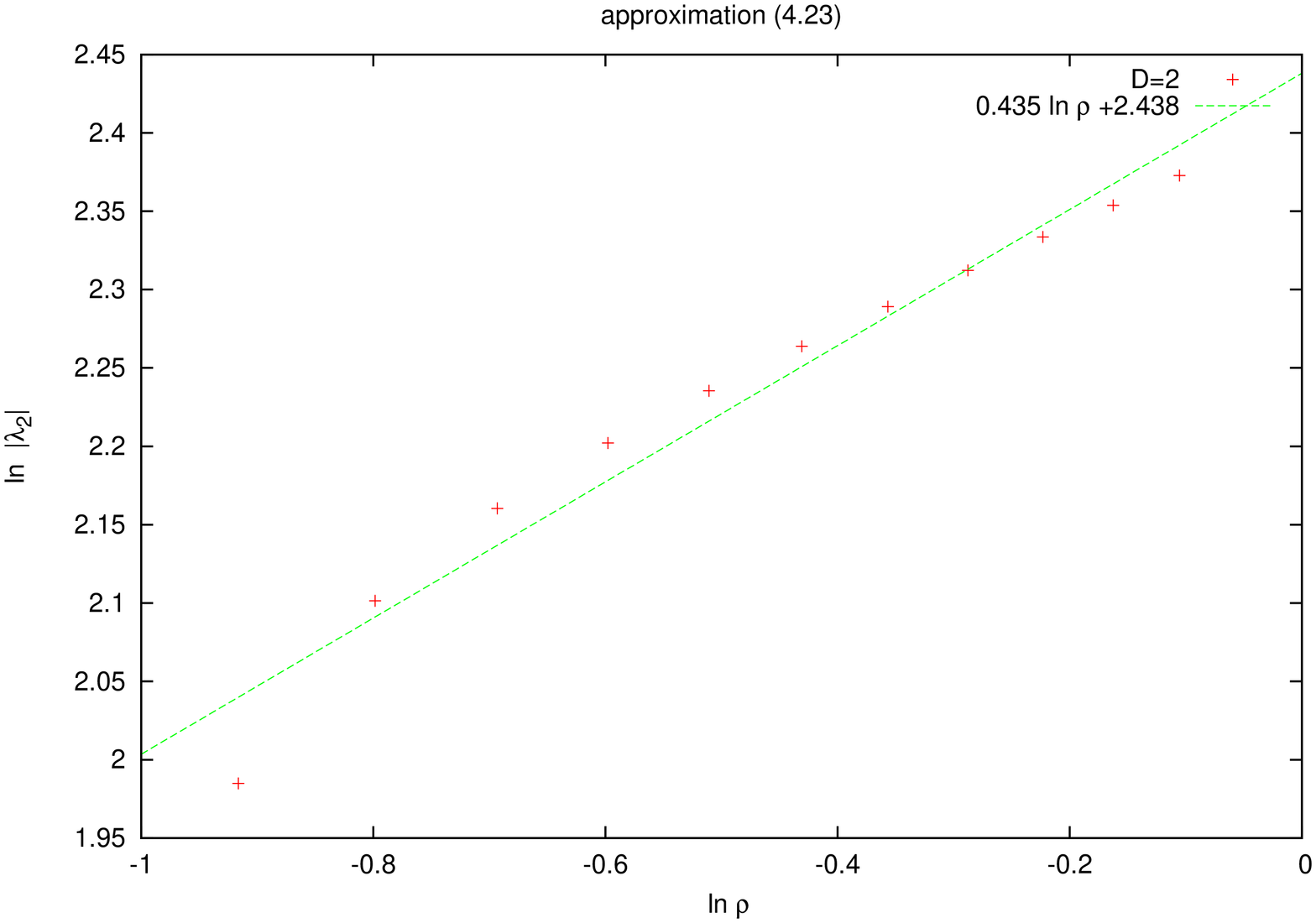}
\caption{The eigenvalues $\lambda_1$ and $\lambda_2$ in the approximations  (\ref{appr3}) in the regime $\ln \rho <-1$. 
}\label{graphs1}
\end{center}
\end{figure}

For $D=2$ the renormalization group steps can be thought of as small and the behavior (\ref{beha}) holds in the regime $\ln \rho < -1$. As it happens this is the most important case corresponding to $d=4$. We find explicitly the following fits

\begin{eqnarray}
\ln |\lambda_1|=-1.296 \ln \rho +0.412~,~\ln |\lambda_2|=0.435 \ln \rho +2.438.
\end{eqnarray}
This result is shown on figure (\ref{graphs1}) together with the results for $\ln |\lambda_1|$ for $D=3$ and $D=4$ and their numerical fits. We also  obtain positive slope for $\ln |\lambda_2|$ for $D=3$ and $D=4$ although the range is much more smaller. We conclude immediately that in all considered dimensions the scaling field $u_1$ corresponding to the mass  is relevant while the scaling field $u_2$ corresponding to the quartic coupling constant is irrelevant. This is the usual conclusion in $d=4$. The critical exponents in $D=2$ are given respectively by
\begin{eqnarray}
y_1=1.296 ~,~y_2=-0.435.
\end{eqnarray}
The mass critical exponent is given by the inverse of the critical exponent $y_1$ associated with the relevant direction, viz $\nu=1/y_1$. The corresponding results for $y_1$ and $\nu$ are included in table (\ref{table3}).
It is very amusing to note that the results shown in table (\ref{table3}) are very close to the average value of $2/d$ and $2/D$, viz
\begin{eqnarray}
\nu=\frac{1}{d}+\frac{1}{D}.
\end{eqnarray}
In the approximation (\ref{appr3}) we obtain instead $\nu=2/D$ in the limit $\rho\longrightarrow 0$ \cite{Nishigaki:1996ts}.

\begin{figure}[htbp]
\begin{center}
\includegraphics[width=5cm,angle=-90]{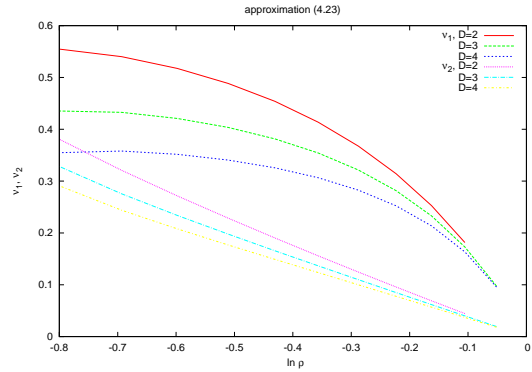}
\caption{The critical exponents  $\nu_1$ and $\nu_2$ in the approximation  (\ref{appr3}) in the regime $\ln \rho <-1$. 
}\label{graphs2}
\end{center}
\end{figure}

An alternative definition of the critical exponents which are associated with the eigenvalues $\lambda_{\alpha}$ is given by  the equation 
 \begin{eqnarray}
\nu_{\alpha}=-\frac{\ln \rho}{\ln |\lambda_{\alpha}|}.
\end{eqnarray}
This is a more naive definition (to our mind) since it does not extract properly the dependence of $\ln |\lambda_{\alpha}|$ on $\ln \rho$ although it is the one that  has been used in \cite{Ferretti:1995zn,Nishigaki:1996ts}. The corresponding results are shown on figure (\ref{graphs2}).

For completeness we have also checked what happens in the approximation  (\ref{appr3}). In this case the eigenvalues $\lambda_1$ and $\lambda_2$ are always real with $\lambda_1>\lambda_2$. The eigenvalue   $\ln |\lambda_{1}|$ as a function of $\ln\rho$ is linear throughout the full regime of $\ln \rho $ whereas $\ln |\lambda_2|$ is linear in the two regimes $\ln \rho <<-1$ and $\ln \rho >-1$ separately. We skip writing down here explicitly the various results for simplicity.

\subsection{The Anomalous Dimension $\eta$}
\begin{figure}[htbp]
\begin{center}
\includegraphics[width=5cm,angle=-90]{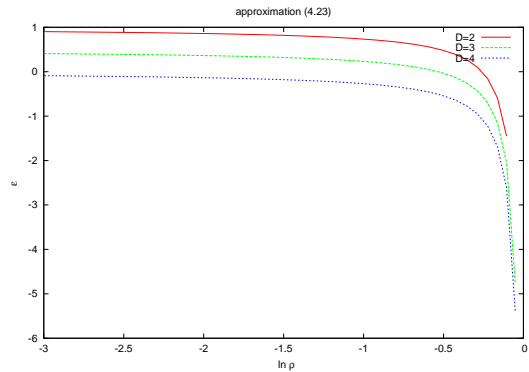}
\caption{The critical exponent  $\eta$ in the approximation  (\ref{appr3}) in the regime $\ln \rho <-1$. 
}\label{graphs3}
\end{center}
\end{figure}

\begin{figure}[htbp]
\centering
\includegraphics[width=15cm,angle=0]{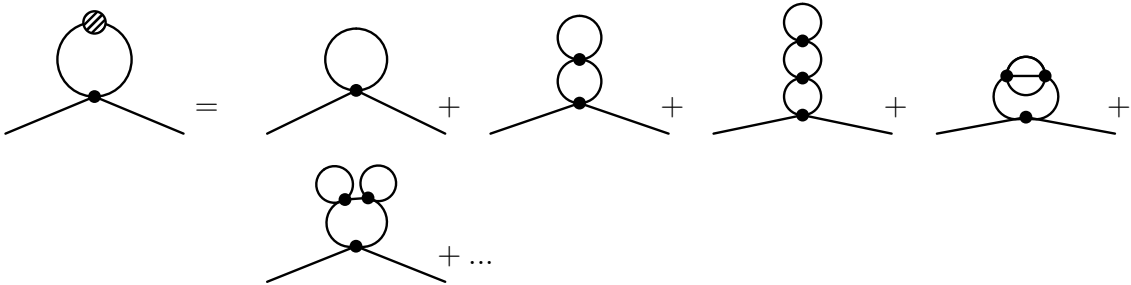}
\caption{Feynman digrams contributing to the renormalization of the mass parameter and the harmonic oscillator coupling constant.}\label{figure1}
\end{figure}

\begin{figure}[htbp]
\centering
\includegraphics[width=15cm,angle=0]{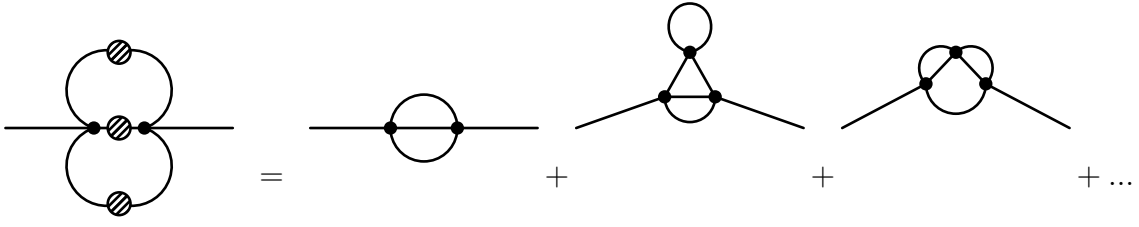}
\caption{Feynman digrams contributing to the wave function renormalization (linear term in $p^2$) and also to the renormalization of the mass parameter and the harmonic oscillator coupling constant (the $p^2=0$ term).}\label{figure2}
\end{figure}

Let us go back to the wave function renormalization contained in the equation

\begin{eqnarray}
\bar{M}^{'}(p)=\rho^{\frac{2+D}{2}}\sqrt{Z(g,\mu^2)+\frac{r^2N}{(c^2+\mu^2)^2}\Delta Z(g,\mu^2)}\bar{M}(\rho p).
\end{eqnarray}
We write this as
\begin{eqnarray}
  \bar{M}(\rho p)=\rho^{-\frac{2+D-\eta}{2}}\bar{M}^{'}(p).
\end{eqnarray}
The coefficient $\eta$ is called the anomalous dimension. It is given by 

\begin{eqnarray}
  \eta=-\frac{\ln\big[Z(g,\mu^2)+\frac{r^2N}{(c^2+\mu^2)^2}\Delta Z(g,\mu^2)\big]}{\ln\rho}.
\end{eqnarray}
At the fixed point we obtain
\begin{eqnarray}
  \eta&=&-\frac{\ln\big[Z(g_{*},\mu_*^2)\big]}{\ln\rho}\nonumber\\
&=&-\frac{\ln(\rho^{-\epsilon}\Gamma_4(g_*)/4g_{*})}{2\ln\rho}\nonumber\\
&=&\frac{\epsilon}{2}-\frac{\ln(\Gamma_4(g_*)/4g_{*})}{2\ln\rho}.
\end{eqnarray}
The anomalous dimension $\eta$ as a function of $\ln \rho$ in the regime $\ln \rho <-1$ in the approximation (\ref{appr3}) is shown on figure (\ref{graphs3}). We observe that $\eta$ approaches a constant value as $\ln \rho \longrightarrow -1$. If we allow ourselves to include also the regime in which the renormalization group eigenvalues $\lambda_1$ and $\lambda_2$ become complex conjugate to each other then we will see that $\eta$ will approach $\epsilon/2$ as $\rho\longrightarrow 0$. This has already been observed in the approximation (\ref{appr4}) in \cite{Ferretti:1995zn,Nishigaki:1996ts}.

\section{The Grosse-Vignes-Tourneret Model}
The renormalizable $\Phi^4$ theory on the degenerate Moyal-Weyl space  ${\bf R}^{d}_{\theta}={\bf R}^D\times {\bf R}^2_{\theta}$ where $D=d-2$ was shown in   \cite{Grosse:2008df} to involve in crucial way the double trace term
\begin{eqnarray}
\frac{\kappa^2}{\theta^2}\nu_2^2\int d^DxTr_{\cal H}\hat{\Phi}^+(x)Tr_{\cal H}\hat{\Phi}(x)=\frac{\kappa^2}{\theta^2}\int d^Dx\int d^2y\int d^2z\Phi^+(x,y)\Phi(x,z).
\end{eqnarray}
In the Landau basis, for example, we can immediately compute  that $\sqrt{\nu}_2Tr_{\cal H}\hat{\Phi}=Tr_H M$ and as a consequence the above term is simply the double trace 
\begin{eqnarray}
\frac{\kappa^2}{\theta^2}\nu_2^2\int d^DxTr_{\cal H}\hat{\Phi}^+(x)Tr_{\cal H}\hat{\Phi}(x)
&=&\frac{2\pi \kappa^2}{\theta}\int d^DxTr_H M^+(x)Tr_H M(x).
\end{eqnarray}
The regularized full self-dual action for a real field becomes then
\begin{eqnarray}
S^{'}[M]&=&\int d^Dx Tr_N\bigg[\frac{1}{2}(\partial_{\mu}M)^2+\frac{1}{2}\mu^2M^2+r^2EM^2+\frac{u}{N}M^4\bigg]+\frac{\pi \kappa^2 r^2}{2}\int d^Dx(Tr_N M)^2.\nonumber\\
\end{eqnarray}
We observe that the new parameter $\kappa^2$ appears always multiplied with the parameter $r^2$ which has important consequences for us later. This theory contains now four independent parameters and as a consequence the renormalization group flow happens in a four dimensional space. 

Now we want to go through the same steps taken in section $4$ and appendix $B$ in order to derive the cumulant expansion. In principle this calculation is quite long but fortunately there is a major simplification due to the simple form of the new added term. This is the first aspect of the calculation which is completely under control.

We expand then the field as $M=\tilde{M}+m$ where $\tilde{M}$ is the slow field and $m$ is the fast field and then we integrate out the fast field $m$. Because of momentum conservation, equation (\ref{B5}), the contribution of the new term to the expansion of the action is trivial. Indeed the mixing term between $\tilde{M}$ and $m$ coming from the new term vanishes due to momentum conservation and as a consequence the expansion  of the action (\ref{B6}) remains unchanged with only the substitution $S\longrightarrow S^{'}$. In particular  after the inclusion of the new term the action $\sigma(m,\tilde{M})$  is still given by equation (\ref{B7})  and as a result the cumulant expansion (\ref{primary}), the fundamental starting point, holds exactly as before.

The last point concerning this aspect of the calculation is how to go from (\ref{primary}) to the cumulant expansion (\ref{mainresult}) which we have used extensively in this article. Originally this step relied on the property that the propagator is diagonal, viz  $<m_{nm}(x)m_{lk}(y)>_0 \sim  \delta_{ml}\delta_{nk}$ which is not the case any longer as we will see shortly. We are therefore left with the only option here which is to work with the more general cumulant expansion (\ref{primary}) which contains four more terms compared to (\ref{mainresult}).

The second aspect that we must consider is the propagator which becomes clearly different with the new added term. In momentum space the quadratic part of the action $S^{'}[m]$ reads (with $\bar{m}(p)$ being the Fourier transform of $m(x)$)
 \begin{eqnarray}
\frac{1}{2}\int \frac{d^Dp}{(2\pi)^D}\bar{m}_{mn}(p)\bigg[\bigg(p^2+\mu^2+r^2(m+n-1)\bigg)\delta_{ml}\delta_{nk}+\pi r^2 \kappa^2\delta_{mn}\delta_{kl}\bigg]\bar{m}^*_{kl}(p).
\end{eqnarray}
 In appendix $C$ we derive an exact formula for the corresponding propagator and show that in the limit $\theta\longrightarrow \infty$ we have 

\begin{eqnarray}
<m_{nm}(x)m_{lk}(y)>_0
&=&\int \frac{d^Dp}{(2\pi)^D}\bigg[\Delta_{mn}(p)\delta_{ml}\delta_{nk}-\pi r^2\kappa^2 \Delta_{mn}(p)\Delta_{kl}(p) \delta_{mn}\delta_{kl}+O(r^4\kappa^4)\bigg] e^{ip(x-y)}.\nonumber\\
\end{eqnarray}
The reduced propagator $\Delta_{mn}(p)$ is given in equation (\ref{appr0}). In this case we do not seem to have simpler analogues of (\ref{appr1}) and  (\ref{appr2}). We stress here that the above equation is really an expansion around $\theta=\infty$ and not an expansion around $\kappa^2=0$ although both are formally equivalent since only the combination $r^2\kappa^2$ which involves the new parameter appear in the original action.

To illustrate the sort of terms which arise in the effective action due to this modification of the propagator we compute the leading corrections, which are proportional to $\kappa$, to the first term of the cumulant expansion  (\ref{primary}).  We find 
\begin{eqnarray}
4\frac{u}{N}\int d^Dx<Tr_N \tilde{M}^2m^2>_{\rm co}&=&...-\pi r^2\kappa^2. 4\frac{u}{N}\int d^Dx \tilde{M}^2(x)_{ii}\bigg[\int_p\Delta^2_{ii}(p)-8\frac{u}{N}\int_{p_1} \int_{p_2}\sum_j\Delta^2_{ij}(p_1)\Delta^2_{jj}(p_2)\nonumber\\
&-&16\frac{u}{N}\int_{p_1} \int_{p_2}\Delta^3_{ii}(p_1)\sum_j\Delta_{ij}(p_2)-4\frac{u}{N}\int_{p_1} \int_{p_2}\Delta_{ii}(p_1)\sum_j\Delta_{jj}(p_1)\Delta_{ij}^2(p_2)\nonumber\\
&-&8\frac{u}{N}\int_{p_1} \int_{p_2}\Delta^2_{ii}(p_1)\sum_j\Delta_{jj}(p_1)\Delta_{ij}(p_2)+O(u^2/N^2)\bigg].
\end{eqnarray}
The dots stands for the terms which were computed in previous sections. By using now the approximations (\ref{appr1}) and  (\ref{appr2}) we obtain
\begin{eqnarray}
4\frac{u}{N}\int d^Dx<Tr_N \tilde{M}^2m^2>_{\rm co}&=&...-\pi r^2\kappa^2. 4\frac{u}{N}\int d^Dx \tilde{M}^2(x)_{ii}\bigg[\int_p\Delta^2_{ii}(p)-8u\int_{p_1} \int_{p_2}\Delta^2_{in_0}(p_1)\Delta^2_{n_0n_0}(p_2)\nonumber\\
&-&16u\int_{p_1} \int_{p_2}\Delta^3_{ii}(p_1)\Delta_{in_0}(p_2)-4u\int_{p_1} \int_{p_2}\Delta_{ii}(p_1)\Delta_{n_0n_0}(p_1)\Delta_{in_0}^2(p_2)\nonumber\\
&-&8u\int_{p_1} \int_{p_2}\Delta^2_{ii}(p_1)\Delta_{n_0n_0}(p_1)\Delta_{in_0}(p_2)+O(u^2/N^2)\bigg].
\end{eqnarray}
These corrections are simply subleading in $1/N$ unless $\kappa^2$ scales with $N$, at least, in such a way that $\kappa^2/N$ is kept fixed. In any case by using now the method of section $4$, for example equation (\ref{pro}), we can immediately conclude that this correction will lead to extra corrections to the mass term $Tr_N\tilde{M}^2$ and the harmonic oscillator term $Tr_NE\tilde{M}^2$ which are proportional to $r^2\kappa^2/N$ and $r^4\kappa^2/N$ respectively. It is obvious, from the index structure, that the term $<Tr_N \tilde{M}^2m^2>_{\rm co}$ can not lead to a correction of the double trace operator $(Tr_N\tilde{M})^2$.

The second term of the cumulant expansion  (\ref{primary}) will obviously lead, among other things, to a correction of the double trace operator $(Tr_N\tilde{M})^2$. The treatment of the third term of  (\ref{primary}), which is the last quadratic term in the background, as well as the treatment of the remaining higher order terms is similar. 

The effective action will therefore involve new extra corrections to the mass term $Tr_N\tilde{M}^2$, to the harmonic oscillator term $r^2Tr_NE\tilde{M}^2$, to the kinetic term $Tr_N(\partial_{\mu}\tilde{M})^2$ and to the quartic interaction  $Tr_N\tilde{M}^4$ proportional to $r^2\kappa^2/N$. It will also involve correction to the double trace operator $(Tr_N\tilde{M})^2$ as expected. The effective action will therefore read (by performing also the second step of the renormalization group procedure which is rescaling of the momenta)

\begin{eqnarray}
 S^{'}[\tilde{M}]+\Delta S^{'}[\tilde{M}]
&=&\frac{\rho^{2+D}}{2}\bigg\{ ...+X_{\rm kin}(g,\mu^2)\frac{r^2\kappa^2}{N} +O(r^4)\bigg\}\int_0^{\Lambda} \frac{d^Dp}{(2\pi)^D} p^2 Tr_N\bar{M}^*(\rho p)\bar{M}(\rho p)\nonumber\\
&+&\frac{\rho^D}{2}\bigg\{...+X_{\rm ma}(g,\mu^2)\frac{r^2\kappa^2}{N}+O(r^4)\bigg\} \int_0^{\Lambda} \frac{d^Dp}{(2\pi)^D}  Tr_N\bar{M}^*(\rho p)\bar{M}(\rho p)\nonumber\\
&+ &  r^2 \rho^D \bigg\{...+X_{\rm h.o}(g,\mu^2)\frac{r^2\kappa^2}{N}+O(r^4)\bigg\} \int_0^{\Lambda} \frac{d^Dp}{(2\pi)^D}  Tr_NE\bar{M}^*(\rho p)\bar{M}(\rho p)\nonumber\\
&+&\frac{u\rho^{3D}}{N}\frac{1}{4g}\bigg\{...+X_{\rm int}(g,\mu^2)\frac{r^2\kappa^2}{N}+O(r^4)\bigg\} \int_0^{\Lambda} \frac{d^Dp_1}{(2\pi)^D}...  \int_0^{\Lambda} \frac{d^Dp_4}{(2\pi)^D}\nonumber\\
&\times &Tr_N\bar{M}(\rho p_1)...\bar{M}(\rho p_4)(2\pi)^D\delta^D(p_1+...p_4)\nonumber\\
&+&\frac{\pi \kappa^2 r^2}{2}\rho^D \bigg\{X_{\kappa}(g,\mu^2)+O(r^2)\bigg\} \int_0^{\Lambda} \frac{d^Dp}{(2\pi)^D}  Tr_N\bar{M}^*(\rho p)Tr_N \bar{M}(\rho p).
\end{eqnarray}
In the above equation $\bar{M}(p)$ is the Fourier transform of $\tilde{M}(x)$ and $g$ is the effective coupling constant replacing $u$ (see table (\ref{table})). The dots stands for the terms computed in previous sections (see equation (\ref{rescaling})). The functions $X_{\rm kin}$, $X_{\rm ma}$, $X_{\rm h.o}$, $X_{\rm int}$ and $X_{\kappa}$ are unknown functions which, as we will see shortly, do not affect the properties of the matrix model fixed point and thus we do not need to determine them explicitly. From the above equation we can deduce immediately the new wave function renormalization to be of the form 

\begin{eqnarray}
\bar{M}^{'}(p)=\rho^{\frac{2+D}{2}}\sqrt{...+X_{\rm kin}(g,\mu^2)\frac{r^2\kappa^2}{N}}\bar{M}(\rho p).
\end{eqnarray}
Let $\tilde{M}^{'}(x)$ be the Fourier transform of $\bar{M}^{'}(p)$. By performing the last step of the renormalization group procedure, which is normalization, we obtain in position space the effective action 
\begin{eqnarray}
 S^{'}[\tilde{M}^{'}]+\Delta S^{'}[\tilde{M}^{'}]&=&\frac{1}{2}\int d^Dx Tr_N(\partial_{\mu}\tilde{M}^{'})^2+\frac{\mu^{'2}}{2}\int d^Dx Tr_N\tilde{M}^{'2}+r^{'2}\int d^Dx Tr_NE \tilde{M}^{'2}\nonumber\\
&+&\frac{u^{'}}{N}\int d^Dx Tr_N\tilde{M}^{'4}+\frac{\pi \kappa^{'2} r^{'2}}{2}\int d^Dx(Tr_N \tilde{M}^{'})^2.
\end{eqnarray}
The renormalized mass ${\mu}^{'2}$, the renormalized quartic coupling constant ${u}^{'}$, the renormalized  inverse noncommutativity $r^{'2}$ and the renormalized parameter $\kappa^{'}$ are given by 
\begin{eqnarray}
{\mu}^{'2}=\frac{\rho^{-2}}{Z}\bigg[...+\frac{r^2\kappa^2}{N}\bigg(X_{\rm ma}-\frac{\Gamma X_{\rm kin}}{Z}\bigg)+O(r^4)\bigg].\label{rg1}
\end{eqnarray}
\begin{eqnarray}
r^{'2}=\frac{\rho^{-2}r^2}{Z}\bigg[...+\frac{r^2\kappa^2}{N}\bigg(X_{\rm h.o}-\frac{\Gamma_e X_{\rm kin}}{Z}\bigg)+O(r^4)\bigg].\label{rg2}
\end{eqnarray}
\begin{eqnarray}
 {u}^{'}=u\frac{\rho^{-\epsilon}}{Z^2(g,\mu^2)}\frac{1}{4g}\bigg[...+\frac{r^2\kappa^2}{N}\bigg(X_{\rm int}-\frac{2\Gamma_4 X_{\rm kin}}{Z}\bigg)+O(r^4)\bigg].\label{rg3} 
\end{eqnarray}
\begin{eqnarray}
 {\kappa}^{'2}=\kappa^2\rho^{D+2}\bigg[\frac{X_k}{\Gamma_e}+O(r^2)\bigg].\label{rg4} 
\end{eqnarray}
From the second renormalization group equation we can see that we still have a fixed point satisfying $r_*=0$ which we identify as the matrix model fixed point. It follows immediately that the renormalization group equations which controls the fixed point are precisely those given by (\ref{mu*}), (\ref{r*}) and (\ref{u*}) and as a consequence the existence and location of the fixed point are unchanged. 

The second very important consequence of the result $r_*=0$ is as follows. The  linearization of the above renormalization group equations (\ref{rg1}), (\ref{rg2}), (\ref{rg3}) and (\ref{rg4}) leads to a renormalization group matrix of the form\footnote{This follows from the linearization of the combination $r^2\kappa^2$ given by $(r^2-r_*^2)\kappa_*^2+r_*^2(\kappa^2-\kappa_*^2)+r_*^2\kappa_*^2$. Thus regardless of the value of the fixed value $\kappa_*^2$ the linearization of $r^2\kappa^2$ is $r^2\kappa_*^2$ since $r_*=0$. }
\begin{eqnarray}
\left( \begin{array}{cccc}
 M_{11} & M_{12} & M_{13} & 0\\
 M_{21} & M_{22} & M_{23} & 0\\
0 & 0 & M_{33} & 0\\
M_{41} & M_{42} & M_{43} & M_{44}
\end{array} \right).
\end{eqnarray}
The upper $3\times 3$ block is given by the original renormalization group matrix  (\ref{rgm}). The corresponding characteristic polynomial gives now the three equations
\begin{eqnarray}
M_{44}-\lambda=0~,~M_{33}-\lambda=0~,~{\rm det} \left( \begin{array}{cc}
 M_{11}-\lambda & M_{12} \\
 M_{21} & M_{22}-\lambda 
\end{array} \right)=0.
\end{eqnarray}
The upper $2\times 2$ block is identical to our previous calculation. Thus the critical exponents $y_1=1/\nu$ and $y_2$ where $\nu$ is the mass critical exponent have the same values found in the original model.

We can also check quite easily that the anomalous dimension $\eta$ has the same value   found in the original model.

In summary by extending the original model to the  Grosse-Vignes-Tourneret model we have convinced ourselves of the following statements:
\begin{itemize}
\item The new extra parameter $\kappa^2$ appears always multiplied by $r^2=4/\theta$. Thus the large $\theta$ limit considered in this article is formally  equivalent to the small $\kappa$ limit.
\item The cumulant expansion is given by (\ref{primary}) and not (\ref{mainresult}) because the extra term in the propagator due to the double trace is not diagonal.
\item The new extra term in the propagator will lead to extra corrections to the mass term $Tr_N\tilde{M}^2$, to the harmonic oscillator term $r^2Tr_NE\tilde{M}^2$, to the kinetic term $Tr_N(\partial_{\mu}\tilde{M})^2$ and to the quartic interaction  $Tr_N\tilde{M}^4$ which in the large $N$ and $\theta$ limits considered in this article are    proportional to $r^2\kappa^2/N$.
\item These corrections are simply subleading in $1/N$ unless $\kappa^2$ scales with $N$, at least, in such a way that $\kappa^2/N$ is kept fixed.
\item There is also correction to the double trace $(Tr_N \tilde{M})^2$.
\item The fixed point and the critical exponents are all unchanged.
\end{itemize}
The conclusion of this section is in accord with the existence of a robust fixed point  which indicates, albeit indirectly, that the model is renormalizable. In fact we can even go further and conclude that the model without the double trace term is renormalizable in the limit considered. A similar conclusion is obtained in  \cite{Becchi:2003dg}. This matrix model fixed point seems to be different from the Ising universality class fixed point\footnote{The $d$ dimensional physics is described  in terms of a critical behavior in $D=d-2$ dimensions.}. A more comprehensive scheme which seems to be more appropriate to describing the matrix transition is to consider instead of the Grosse-Vignes-Tourneret model the most general quartic action containing all multi trace operators consistent with the symmetry $M\longrightarrow -M$. Thus the extra terms $(Tr_N M^2)^2$, $(Tr_N M)^4$ and $(Tr_N M)^2 Tr_N M^2$ should also be included. For example the term $(Tr_N M^2)^2$ was found to play a crucial role in the renormalizability of noncommutative phi-four theory in the large $\theta$ limit in \cite{Becchi:2003dg}. 
We hope to return to this point as well as other points (see towards the end of next section) in the future.  

\section{Conclusion and Outlook}
In this article we have presented a study of phi-four theory on noncommutative spaces with only two noncommuting directions at the self-dual point using a combination of the  Wilson  approximate renormalization group recursion formula and the solution to the corresponding  zero dimensional matrix model at large $N$. The most important results of this work are:
\begin{itemize}
\item The Penner matrix model $Tr_N (M^2/2+ m^2 EM^2 +gM^4/N)$ can be systematically solved by the multi trace approach of \cite{O'Connor:2007ea} as illustrated by the computation of the $2-$point proper vertex in section $3$. This might be an alternative approach to the one pursued in \cite{Grosse:2012uv}. 
\item The action studied in this article is a generalization of the Penner matrix model of the form
\begin{eqnarray}
S[M]&=&\int d^Dx Tr_N\bigg[\frac{1}{2}(\partial_{\mu}M)^2+\frac{1}{2}\mu^2M^2+r^2EM^2+\frac{u}{N}M^4\bigg].
\end{eqnarray}
\item The renormalizations of the mass term, the harmonic oscillator term and the quartic interaction are straightforward to obtain in this scheme. 
\item  The most difficult calculation of all is wave function renormalization. We have conjectured in this article that wave function renormalization within the scheme of the recursion formula is fully encoded in the approximation (\ref{appr3}).  
\item We have found a fixed point solution of the renormalization group equations (\ref{rgf1}), (\ref{rgf2}) and (\ref{rgf3}) with $r_*^2=0$, $\mu_*^2<0$ and $u_*>0$ in all dimensions $D=2,3,4$ corresponding to the dimensions $d=4,5,6$  respectively. We do not exclude here the possibility of the existence of other fixed point solutions even within this scheme.
\item The renormalization group eigenvalue $\lambda_3$ was determine to be larger than $1$ and hence $r^2$ is a relevant coupling similar to the mass. This statement was however derived  from the available perturbative knowledge of the proper vertex $\Gamma_e$ and thus should be taken with care. 
\item We have found that the renormalization group eigenvalues $\lambda_1$ (corresponding to the mass, relevant) and $\lambda_2$ (corresponding to the quartic coupling, irrelevant) become complex conjugate of each other at the value of the dilatation parameter $\ln \rho \simeq -1$. After extracting the dependence on $\rho$ we obtain the mass critical exponent 
\begin{eqnarray}
\nu\simeq \frac{1}{d}+\frac{1}{D}.
\end{eqnarray}
In $D=2$ (the most important case for us) it is found that both the eigenvalues $\ln \lambda_1$ and $\ln \lambda_2$ scale linearly with $\ln \rho$ over the entire real range $\ln \rho< -1$. We obtain the indices 
\begin{eqnarray}
y_1=1.296 ~(\nu=0.772)~,~y_2=-0.435.
\end{eqnarray}
\item The anomalous dimension was found to scale as
\begin{eqnarray}
\eta\simeq \frac{4-D}{2}~,~\rho \rightarrow 0
\end{eqnarray}
\item Extension of the above results to the  Grosse-Vignes-Tourneret model, which involves an extra term given by the double trace operator $(Tr_N M)^2$, was also given.
\end{itemize}
We conclude this article by indicating that the most natural extension of this work should be to repeat the same analysis of the matrix model fixed point by replacing  the recursion formula with the  exact functional renormalization group method. The functional renormalization group (as opposed to other forms of the renormalization group) is the most direct implementation of the Wilson idea which goes along the lines of the recursion formula but is exact although explicit calculation will undoubtedly   involve various truncations which by their nature are also approximate.

\paragraph{Acknowledgments:} This research was supported by “The National Agency for the
Development of University Research (ANDRU)” under PNR contract number U23/Av58
(8/u23/2723).

\begin{table}[h]
\centering
\resizebox{15cm}{!}{
\begin{tabular}{|l|c|c| }
\hline
function & non-perturbative  & perturbative\\
\hline 
$g$ & $v_Du(c^2+\mu^2+r^2N)^{-2}$ & \\
\hline
$h=1/\kappa $ & $(c^2+\mu^2+r^2N)(c^2+\mu^2+r^2N/2)^{-1}$ & $1+r^2N/(2(c^2+\mu^2))+...$\\
\hline 
$a^2$ & $(\sqrt{1+48g}-1)/{24g}$ & $1-12g +288 g^2+...$\\
\hline 
$z$ & $(\mu^2-\epsilon_0 c^2)/(c^2+\mu^2)$ & \\
\hline
$Z_2(g)\equiv Z_2(g,1)$ & 
$((1-a^2)^2(5-2a^2))/(18a^2(4-a^2)) $
 & $8g^2-256g^3+...$ \\
\hline
$Z(g,\mu^2)$ & $1+2 zZ_2(g) $ & \\
\hline
$\Gamma_2(g)\equiv \Gamma_2(g,1)+1$ & 
${3}/(a^2(4-a^2)) $
 & $1+8g-80g^2+1664g^3-...$ \\
\hline
$\Gamma(g,\mu^2)$ & $(\mu^2+c^2)\Gamma_2(g)-c^2$ &\\
\hline
$\Gamma_4(g)=\Gamma_4(g,1)$ & 
$(9(1-a^2)(5-2a^2))/(a^4(4-a^2)^4) $
 & $4g-32g^2+896g^3+...$ \\
\hline
$Z_2(g,h)$ & 
$?$
 & $8g^2h^3-256g^3h^4+...$ \\
\hline
$\Gamma_2(g,h)$ & 
$?$
 & $8gh-80g^2h^2+2g^3 (256h^2+576h^3)+...$ \\
\hline
$\Gamma_4(g,h)$ & 
$?$
 & $4g(1-8gh^2+g^2(160h^4+64h^3)+...)$ \\
\hline
$\Gamma_e(g,h)$ & $1+\partial_{\kappa}\Gamma_2(g,h)/2$ & \\
\hline
$\Gamma_e(g)$ & $\Gamma_e(g,1)$ & \\
\hline
$\Delta Z(g,\mu^2)$ & 
$ c^2(1+\epsilon_0\rho^2)Z_2(g)+({\mu}^{2}-\epsilon_0\rho^2 c^2)\partial_hZ_2(g,h)|_{h=1} $
 &  \\
\hline
$\Delta \Gamma(g)$ & 
$ \Gamma_2(g)-\Gamma_e(g) $
 &  \\
\hline
$\Delta\Gamma_4(g)$ & 
$ \partial_h{\Gamma}_4(g,h)|_{h=1}/2$
 & \\
\hline
$\Delta \Gamma_e(g)$ & $-\partial_{\kappa}\Gamma_e(g,h)|_{\kappa=1}/2$ & \\
\hline
\end{tabular}
}
\caption{The different functions appearing in the effective action (\ref{effectiveactionSD2}). }\label{table}
\end{table}

\begin{table}[h]
\centering
\resizebox{5cm}{!}{
\begin{tabular}{|l|c|c|c| }
\hline
$D (d)$ & $g_{*}$  & ${\mu}_*^2$ & ${u}_*$\\
\hline 
$2(4)$ & $1.845$ & $-0.693$ & $2.92$\\
\hline 
$3(5)$ & $2.235$ & $-0.732$ & $10.853$\\
\hline 
$4(6)$ & $2.518$ & $-0.757$ & $50.083$\\
\hline
\end{tabular}
}
\caption{The critical values for $\rho=0.5$ in the approximation (\ref{appr3}). }\label{table1}
\end{table}

\begin{table}[h]
\centering
\resizebox{5cm}{!}{
\begin{tabular}{|l|c|c|c| }
\hline
$D (d)$ & $g_{*}$  & ${\mu}_*^2$ & ${u}_*$\\
\hline 
$2(4)$ & $2.282$ & $-0.851$ & $0.854$\\
\hline 
$3(5)$ & $0.409$ & $-0.643$ & $3.527$\\
\hline 
$4(6)$ & $0$ & $0$ & $0$\\
\hline
\end{tabular}
}
\caption{The critical values for $\rho=0.5$ in the approximation  (\ref{appr4}). }\label{table2}
\end{table}

\begin{table}[h]
\centering
\resizebox{4cm}{!}{
\begin{tabular}{|l|c|c| }
\hline
$D (d)$ & $y_1$  & ${\nu}$ \\
\hline 
$2(4)$ & $1.296$ & $0.772$ \\
\hline 
$3(5)$ & $1.851$ & $0.540$ \\
\hline 
$4(6)$ & $2.345$ & $0.426$ \\
\hline
\end{tabular}
}
\caption{The critical exponents $y_1$ and  $\nu$ in the approximation (\ref{appr3}). }\label{table3}
\end{table}

\appendix
\section{Degenerate Duality Transformations}
Let us start with the quadratic action. We have (with $D_i=\partial_i-iB_{ij}x_j$ and $C_i=\partial_i+iB_{ij}x_j$)
\begin{eqnarray}
S_2[\Phi,B]&=&\int
d^dx~ \Phi^+\bigg(-\frac{1}{2}{\partial}_i^2+\frac{1}{2}(B_{ij}x_j)^2-\frac{1}{2}\partial_{\mu}^2+\frac{{\mu}^2}{2}\bigg)\Phi\nonumber\\
&=&\int
d^Dx\int d^2x~ \bigg(-\frac{1}{4}(D_i\Phi)^+(D_i\Phi)-\frac{1}{4}(C_i\Phi)^+(C_i\Phi)-\frac{1}{2}\Phi^+\partial_{\mu}^2\Phi^++\frac{{\mu}^2}{2}\Phi^+\Phi\bigg)\nonumber\\
&=&\int\frac{d^Dp}{(2\pi)^D}\int d^2x~ \bigg(-\frac{1}{4}(D_i\Phi)^+(p,x)(D_i\Phi)(p,x)-\frac{1}{4}(C_i\Phi)^+(p,x)(C_i\Phi)(p,x)\nonumber\\
&+&\frac{1}{2}(p^2+{\mu}^2)\Phi^+(p,x)\Phi(p,x)\bigg).
\end{eqnarray}
We have defined Fourier transform in the commuting directions by

\begin{eqnarray}
f(x^{\mu},x^i)=\int \frac{d^Dp}{(2\pi)^D} f(p^{\mu},x^i)~e^{ip^{\mu}x_{\mu}}.
\end{eqnarray}
We define $\tilde{k}_i=B^{-1}_{ij}k_j$. The Fourier transform of $\Phi(p,x)$ and $D_i\Phi(p,x)$ in the noncommuting directions are $\tilde{\Phi}(p,k)$ and $-\tilde{D}_i\tilde{\Phi}(p,k)$ where 
\begin{eqnarray}
\tilde{\Phi}(p,k)=\int d^2x \Phi(p,x)~e^{-ik^ix_i}.
\end{eqnarray}
\begin{eqnarray}
-\tilde{D}_i\tilde{\Phi}(p,k)&=&\int d^2x D_i\Phi(p,x)~e^{-ik^ix_i}\nonumber\\
&=&\frac{\partial}{\partial\tilde{k}_i}-iB_{ij}\tilde{k}_j.
\end{eqnarray}
We can then compute immediately that 
\begin{eqnarray}
\int\frac{d^Dp}{(2\pi)^D}\int d^2x~ (D_i\Phi)^+(p,x)(D_i\Phi)(p,x)&=&\int\frac{d^Dp}{(2\pi)^D}\int d^2\tilde{k}~ (\tilde{D_i}\hat{\Phi})^+(p,\tilde{k})(\tilde{D}_i\hat{\Phi})(p,\tilde{k}).\nonumber\\
\end{eqnarray}
The new field $\hat{\Phi}$ is defined by
\begin{eqnarray}
\hat{\Phi}(p,\tilde{k})=\sqrt{|{\rm det}\frac{B}{2\pi}|}\tilde{\Phi}(p,B\tilde{k}).
\end{eqnarray}
A similar result holds for the other quadratic terms. By renaming the variable as $\tilde{k}=x$ we can see that the resulting quadratic action has therefore the same form as the original quadratic action, viz 
\begin{eqnarray}
S_2[\Phi,B]=S_2[\hat{\Phi},B].
\end{eqnarray}
Next we consider the interaction term 
\begin{eqnarray}
S_{\rm int}[\Phi,B]&=&\int d^dx \Phi^+*\Phi *\Phi^+ *\Phi\nonumber\\
&=&\int_{p_1}...\int_{p_4}(2\pi)^D\delta^D(p_1-p_2+p_3-p_4)\int d^2x \Phi^+(p_1,x^i)*\Phi(p_2,x^i)*\Phi^+(p_3,x^i)*\Phi^+(p_4,x^i)\nonumber\\
&=&\int_{p_1}...\int_{p_4}(2\pi)^D\delta^D(p_1-p_2+p_3-p_4)\int_{k_1}...\int_{k_4} \tilde{\Phi}^+(p_1,k_1)\tilde{\Phi}(p_2,k_2)\tilde{\Phi}^+(p_3,k_3)\tilde{\Phi}^+(p_4,k_4)\nonumber\\
&\times &\tilde{V}(k_1,k_2,k_3,k_4).
\end{eqnarray}
The vertex in momentum space is given by
\begin{eqnarray}
\tilde{V}(k_1,k_2,k_3,k_4)=(2\pi)^2\delta^2(k_1-k_2+k_3-k_4)e^{-i\theta^{\mu\nu}\big(k_{1\mu}k_{2\nu}+k_{3\mu}k_{4\nu}\big)}.
\end{eqnarray}
By substituting $k=B\tilde{k}$ we obtain 
\begin{eqnarray}
S_{\rm int}[\Phi,B]
&=&\int_{p_1}...\int_{p_4}(2\pi)^D\delta^D(p_1-p_2+p_3-p_4)\int_{\tilde{k}_1}...\int_{\tilde{k}_4} \hat{\Phi}^+(p_1,\tilde{k}_1)\hat{\Phi}(p_2,\tilde{k}_2)\hat{\Phi}^+(p_3,\tilde{k}_3)\hat{\Phi}^+(p_4,\tilde{k}_4)\nonumber\\
&\times &\hat{V}(k_1,k_2,k_3,k_4).\label{int1}
\end{eqnarray}
The new vertex is given by
\begin{eqnarray}
\hat{V}(\tilde{k}_1,\tilde{k}_2,\tilde{k}_3,\tilde{k}_4)=\frac{{\rm det}{B}}{(2\pi)^2} \delta^2(\tilde{k}_1-\tilde{k}_2+\tilde{k}_3-\tilde{k}_4)e^{i(B{\theta}B)^{\mu\nu}\big(\tilde{k}_{1\mu}\tilde{k}_{2\nu}+\tilde{k}_{3\mu}\tilde{k}_{4\nu}\big)}.\label{int2}
\end{eqnarray}
The interaction term can also be rewritten as
\begin{eqnarray}
S_{\rm int}[\Phi,B]
&=&\int_{p_1}...\int_{p_4}(2\pi)^D\delta^D(p_1-p_2+p_3-p_4)\int_{x_1}...\int_{x_4} {\Phi}^+(p_1,x_1){\Phi}(p_2,x_2){\Phi}^+(p_3,x_3){\Phi}^+(p_4,x_4)\nonumber\\
&\times &{V}(x_1,x_2,x_3,x_4).\label{int3}
\end{eqnarray}
The vertex in position space is given by
\begin{eqnarray}
{V}(x_1,x_2,x_3,x_4)&=&\int_{k_1}...\int_{k_4}\tilde{V}(k_1,k_2,k_3,k_4)e^{ik_1x_1-ik_2x_2+ik_3x_3-ik_4x_4}\nonumber\\
&=&\frac{1}{(2\pi)^2|{\rm det}\theta|}\delta^2(x_1-x_2+x_3-x_4)e^{-i(\theta^{-1})_{\mu\nu}(x_1^{\mu}x_2^{\mu}+x_3^{\mu}x_4^{\mu})}.\label{int4}
\end{eqnarray}
We can see immediately from comparing equations (\ref{int1}),(\ref{int2}) and equations (\ref{int3}),(\ref{int4}) that the interaction term in momentum space has the same form as the interaction term in position space provided that the new noncommutativity parameter and the new coupling constant are given by 
\begin{eqnarray}
\hat{\theta}=-B^{-1}\theta^{-1}B^{-1}.
\end{eqnarray}
\begin{eqnarray}
\lambda \frac{{\rm det}B}{(2\pi)^2}=\frac{\hat{\lambda}}{(2\pi)^2}\frac{1}{{\rm det}\hat{\theta}}\Leftrightarrow \hat{\lambda}=\frac{\lambda}{|{\rm det}B\theta |}.
\end{eqnarray}

\section{Detailed Calculation of (\ref{mainresult})}

The action of interest is

\begin{eqnarray}
S[M]
&=&\nu_2\int d^{D}x~\sum_{m,n,k,l}\bigg(\frac{1}{2}(M^+)_{mn}G_{mn,kl}M_{kl}+\frac{\lambda }{4!}(M^+)_{mn}M_{nk}(M^+)_{kl}M_{lm}\bigg).\nonumber\\
\end{eqnarray}
For the case $\omega=0$ the Laplacian takes the form 
\begin{eqnarray}
G_{mn,kl}&=&\big(-{\partial}_{\mu}^2+{\mu}^2+ {\mu}_1^2(m+n-1)\big){\delta}_{n,k}{\delta}_{m,l}.
\end{eqnarray}
We will consider only the case of hermitian matrices $M=M^+$. Furthermore we will apply the renormalization group method as outline in the seminal work \cite{Wilson:1973jj} which goes also under the name of the recursion formula.

We decompose  the $N\times N$ matrix $M$ into an $N\times N$ background matrix $\tilde{M}$ and an $N\times N$ fluctuation matrix $m$, viz
\begin{eqnarray}
M=\tilde{M}+m.
\end{eqnarray}
We introduce a cutoff $\Lambda$. The background $\tilde{M}$ contains slow modes, i.e. modes with momenta less or equal than $\rho\Lambda$ while the fluctuation $m$  contains fast modes, i.e. modes with momenta  larger  than $\rho\Lambda$ where $0<\rho<1$. Explicitly we have
\begin{eqnarray}
\tilde{M}_{mn}(x)=\int_{0\leq|p|\leq\rho\Lambda} \frac{d^{D}p}{(2\pi)^D}~\tilde{M}_{mn}(p)~e^{ipx}~,~m_{mn}(x)=\int_{\rho\Lambda<|p|<\Lambda} \frac{d^{D}p}{(2\pi)^D}~{m}_{mn}(p)~e^{ipx}.\nonumber\\
\end{eqnarray}
Clearly
\begin{eqnarray}
\int d^Dx \tilde{M}_{mn}(x)m_{kl}(x)=\int d^Dx \tilde{M}_{mn}(x)(-\partial_{\mu}^2)m_{kl}(x)=0.\label{B5}
\end{eqnarray}
This is an expression of momentum conservation. Next we compute
\begin{eqnarray}
S[M]=S[m]+S[\tilde{M}]+\sigma(m,\tilde{M}).\label{B6}
\end{eqnarray}
\begin{eqnarray}
\sigma(m,\tilde{M})
&=&\nu_2\frac{\lambda}{4!}\int d^Dx Tr(2\tilde{M}m\tilde{M}m+4\tilde{M}^2m^2+4\tilde{M}^3m+4\tilde{M}m^3).\label{B7}
\end{eqnarray}
 We are interested in the partition function 
\begin{eqnarray}
Z=\int dM e^{-S[M]}&=&\int d\tilde{M}e^{-S[\tilde{M}]}\int dm e^{-S[m]}e^{-\sigma(m,\tilde{M})}\nonumber\\
&=&\int d\tilde {M}e^{-S[\tilde{M}]}<e^{-\sigma(m,\tilde{M})}>\int dm e^{-S[m]}.
\end{eqnarray}
We have defined
\begin{eqnarray}
<{\cal O}>=\frac{\int dm {\cal O} e^{-S[m]}}{\int dm e^{-S[m]}}.
\end{eqnarray}
The free propagator is
\begin{eqnarray}
<m_{nm}(x)m_{lk}(y)>_0
&=&\frac{1}{\nu_2}\Delta_{mn,kl}(x-y)\nonumber\\
&=&\frac{1}{\nu_2}\int \frac{d^Dp}{(2\pi)^D}\frac{\delta_{n,k}\delta_{m,l}}{p^2+\mu^2+\mu_1^2(m+n-1)}e^{ip(x-y)}.
\end{eqnarray}
Clearly we can write
\begin{eqnarray}
<e^{-\sigma(m,\tilde{M})}>&=&e^{-\Delta S_{\rm eff}[\tilde{M}]}.
\end{eqnarray}
We compute up to the $4$th order in the slow matrix $\tilde{M}$ the non perturbative expansion (with $u=\nu_2\lambda/4!$)
\begin{eqnarray}
<e^{-\sigma(m,\tilde{M})}>&=&<\bigg[1-u\bigg({\cal M}_1+{\cal M}_2+{\cal M}_3\bigg)+\frac{u^2}{2}\bigg({\cal M}_1^2+{\cal M}_2^2+2{\cal M}_1{\cal M}_2+2{\cal M}_1{\cal M}_3\bigg)\nonumber\\
&-&\frac{u^3}{6}\bigg({\cal M}_1^3+3{\cal M}_1^2{\cal M}_2\bigg)+\frac{u^4}{24}{\cal M}_1^4+...\bigg]>.
\end{eqnarray}
\begin{eqnarray}
{\cal M}_1=4\int d^D x Tr\tilde{M}m^3~,~{\cal M}_2=2\int d^Dx Tr\tilde{M}m\tilde{M}m+4\int d^Dx Tr\tilde{M}^2m^2~,~{\cal M}_3=4\int d^Dx Tr\tilde{M}^3m.\nonumber\\
\end{eqnarray}
By using the symmetry $m\longrightarrow -m$ we observe that the terms $<{\cal M}_1>$, $<{\cal M}_3>$, $<{\cal M}_1{\cal M}_2>$ and ${\cal M}_1^3$ vanish. We then get
\begin{eqnarray}
<e^{-\sigma(m,{\cal M})}>&=&<\bigg[1-u{\cal M}_2+\frac{u^2}{2}\bigg({\cal M}_1^2+{\cal M}_2^2+2{\cal M}_1{\cal M}_3\bigg)-\frac{u^3}{2}{\cal M}_1^2{\cal M}_2+\frac{u^4}{24}{\cal M}_1^4+...\bigg]>.\label{cumu}\nonumber\\
\end{eqnarray}
We want to calculate
\begin{eqnarray}
<{\cal M}_1{\cal M}_3>=16\int d^Dx \int d^Dy\tilde{M}(x)_{ij}\tilde{M}^3(y)_{kl}<m^3(x)_{ji}m(y)_{lk}>.
\end{eqnarray}
This vanishes by momentum conservation since the momentum $p$ carried by the field $\tilde{M}_{ij}$ at point $x$ will be transferred to the field $m_{lk}$ at point $y$ after interacting via a $4-$point vertex.

We redefine the  path integral over the fluctuation fields $m$ by
\begin{eqnarray}
Z[J,b]&=&Z[J,b_1,b_2,b_{1+1},b_3]\nonumber\\
&=&\int dm~\exp\bigg[-S[m]-\int d^dx \bigg(Jm+b_1Tr\tilde{M}^3m+b_2Tr\tilde{M}^2m^2+b_{1+1}Tr\tilde{M}m\tilde{M}m+b_3Tr\tilde{M}m^3\bigg)\bigg].\nonumber\\
\end{eqnarray}
The corresponding vacuum energy is defined by
\begin{eqnarray}
Z[J,b]=\exp(-W[J,b]).
\end{eqnarray}
We will need among other things the following connected Green's functions 
\begin{eqnarray}
<Tr\tilde{M}m^3(x_1)...Tr\tilde{M}m^3(x_n)>_{\rm co}&=&(-1)^{n+1}\frac{1}{\nu_2^n}\frac{\delta^n W[J,b]}{\delta b_3(x_1)...\delta b_3(x_n)}|_{J=b=0}.
\end{eqnarray}
In fact we want to calculate expectation values of products of $U(N)-$singlets such as $Tr\tilde{M}m^3$, $Tr\tilde{M}^3m$, $Tr\tilde{M}^2m^2$ and $Tr\tilde{M}m\tilde{M}m$ (both $m$ and $\tilde{M}$ transform under $U(N)$ transformations in the same way).  We can safely assume that connected Green's functions with an odd number of external legs vanish. 

In the remainder of this appendix we proceed to the explicit evaluation of the cumulant expansion  (\ref{cumu}). First we compute
\begin{eqnarray}
<{\cal M}_1^4>&=&256\int d^Dx_1\int d^Dx_2\int d^Dx_3\int d^Dx_4<Tr\tilde{M}m^3(x_1)...Tr\tilde{M}m^3(x_4)>\nonumber\\
&=&256\int d^Dx_1...\int d^Dx_4\frac{1}{Z[J,b]\nu_2^4}\frac{\delta^4}{\delta b_3(x_1)\delta b_3(x_2)\delta b_3 (x_3) \delta b_3(x_4)}Z[J,b]|_{J=b=0}\nonumber\\
&=&256\times 3 \bigg(\int d^Dx_1\int d^Dx_2<Tr\tilde{M}m^3(x_1)Tr\tilde{M}m^3(x_2)>_{\rm co}\bigg)^2\nonumber\\
&+&256\int d^Dx_1...\int d^Dx_4<Tr\tilde{M}m^3(x_1)...Tr\tilde{M}m^3(x_4)>_{\rm co}.
\end{eqnarray}
The factor of $3$ in the third line comes from the three disconnected graphs contributing to the $4-$point function. 

Similarly
\begin{eqnarray}
<{\cal M}_1^2>&=&16\int d^Dx\int d^Dy<Tr\tilde{M}m^3(x)Tr\tilde{M}m^3(y)>\nonumber\\
&=&16\int d^Dx\int d^Dy<Tr\tilde{M}m^3(x)Tr\tilde{M}m^3(y)>_{\rm co}.
\end{eqnarray}
Next we compute
\begin{eqnarray}
<{\cal M}_1^2{\cal M}_2>&=&64\int d^Dx_1\int d^Dx_2\int d^Dx_3<Tr\tilde{M}m^3(x_1)Tr\tilde{M}m^3(x_2)Tr\tilde{M}^2m^2(x_3)>\nonumber\\
&+&32\int d^Dx_1\int d^Dx_2\int d^Dx_3<Tr\tilde{M}m^3(x_1)Tr\tilde{M}m^3(x_2)Tr\tilde{M}m\tilde{M}m(x_3)>.\nonumber\\
\end{eqnarray}
First we have
\begin{eqnarray}
<Tr\tilde{M}m^3(x_1)Tr\tilde{M}m^3(x_2)Tr\tilde{M}^2m^2(x_3)>&=&-\frac{1}{Z[J,b]\nu_2^3}\frac{\delta^3}{\delta b_3(x_1)\delta b_3(x_2)\delta b_2 (x_3)}Z[J,b]|_{J=b=0}\nonumber\\
&=&\frac{1}{\nu_2^3}\frac{\delta^3}{\delta b_3(x_1)\delta b_3(x_2)\delta b_2 (x_3)}W[J,b]|_{J=b=0}\nonumber\\
&-&\frac{1}{\nu_2^3}\frac{\delta^2}{\delta b_3(x_1)\delta b_3(x_2)}W[J,b]|_{J=b=0}\frac{\delta}{\delta b_2 (x_3)}W[J,b]|_{J=b=0}.\nonumber\\
&=&<Tr\tilde{M}m^3(x_1)Tr\tilde{M}m^3(x_2)Tr\tilde{M}^2m^2(x_3)>_{\rm co}\nonumber\\
&+&<Tr\tilde{M}m^3(x_1)Tr\tilde{M}m^3(x_2)>_{\rm co}<Tr\tilde{M}^2m^2(x_3)>_{\rm co}.\nonumber\\
\end{eqnarray}
We also have
\begin{eqnarray}
<Tr\tilde{M}m^3(x_1)Tr\tilde{M}m^3(x_2)Tr\tilde{M}m\tilde{M}m(x_3)>&=&-\frac{1}{Z[J,b]\nu_2^3}\frac{\delta^3}{\delta b_3(x_1)\delta b_3(x_2)\delta b_{1+1} (x_3)}Z[J,b]|_{J=b=0}\nonumber\\
&=&\frac{1}{\nu_2^3}\frac{\delta^3}{\delta b_3(x_1)\delta b_3(x_2)\delta b_{1+1} (x_3)}W[J,b]|_{J=b=0}\nonumber\\
&-&\frac{1}{\nu_2^3}\frac{\delta^2}{\delta b_3(x_1)\delta b_3(x_2)}W[J,b]|_{J=b=0}\frac{\delta}{\delta b_{1+1} (x_3)}W[J,b]|_{J=b=0}.\nonumber\\
&=&<Tr\tilde{M}m^3(x_1)Tr\tilde{M}m^3(x_2)Tr\tilde{M}^2m^2(x_3)>_{\rm co}\nonumber\\
&+&<Tr\tilde{M}m^3(x_1)Tr\tilde{M}m^3(x_2)>_{\rm co}<Tr\tilde{M}m\tilde{M}m(x_3)>_{\rm co}.\nonumber\\
\end{eqnarray}
The definition of the connected Green's functions $<Tr\tilde{M}m^3(x_1)Tr\tilde{M}m^3(x_2)Tr\tilde{M}^2m^2(x_3)>_{\rm co}$, $<Tr\tilde{M}m^3(x_1)Tr\tilde{M}m^3(x_2)Tr\tilde{M}m\tilde{M}m(x_3)>_{\rm co}$, $<Tr\tilde{M}^2m^2(x_3)>_{\rm co}$ and $<Tr\tilde{M}m\tilde{M}m(x_3)>_{\rm co}$ is obvious from the above equations.

Also we compute

\begin{eqnarray}
<{\cal M}_2^2>&=&4\int d^Dx\int d^Dy<Tr\tilde{M}m\tilde{M}m(x)Tr\tilde{M}m\tilde{M}m(y)>\nonumber\\
&+&16\int d^Dx\int d^Dy<Tr\tilde{M}^2m^2(x)Tr\tilde{M}^2m^2(y)>\nonumber\\
&+&16\int d^Dx\int d^Dy<Tr\tilde{M}m\tilde{M}m(x)Tr\tilde{M}^2m^2(y)>\nonumber\\
&=&4\int d^Dx\int d^Dy <Tr\tilde{M}m\tilde{M}m(x)Tr\tilde{M}m\tilde{M}m(y)>_{\rm co}+4\bigg(\int d^Dx <Tr\tilde{M}m\tilde{M}m(x)>_{\rm co}\bigg)^2\nonumber\\
&+&16\int d^Dx\int d^Dy <Tr\tilde{M}^2m^2(x)Tr\tilde{M}^2m^2(y)>_{\rm co}+16\bigg(\int d^Dx <Tr\tilde{M}^2m^2(x)>_{\rm co}\bigg)^2\nonumber\\
&+&16\int d^Dx\int d^Dy<Tr\tilde{M}m\tilde{M}m(x)Tr\tilde{M}^2m^2(y)>_{\rm co}\nonumber\\
&+&16\int d^Dx <Tr\tilde{M}m\tilde{M}m(x)>_{\rm co}\int d^Dy <Tr\tilde{M}^2m^2(y)>_{\rm co}.
\end{eqnarray}
We have defined the connected two-point functions
\begin{eqnarray}
<Tr\tilde{M}^2m^2(x)Tr\tilde{M}^2m^2(y)>_{\rm co}&=&-\frac{1}{\nu_2^2}\frac{\delta^2 W[J,b]}{\delta b_2(x)\delta b_2(y)}|_{J=b=0}.
\end{eqnarray}
\begin{eqnarray}
<Tr\tilde{M}m\tilde{M}m(x)Tr\tilde{M}m\tilde{M}m(y)>_{\rm co}&=&-\frac{1}{\nu_2^2}\frac{\delta^2 W[J,b]}{\delta b_{1+1}(x)\delta b_{1+1}(y)}|_{J=b=0}.
\end{eqnarray}
\begin{eqnarray}
<Tr\tilde{M}m\tilde{M}m(x)Tr\tilde{M}^2m^2(y)>_{\rm co}&=&-\frac{1}{\nu_2^2}\frac{\delta^2 W[J,b]}{\delta b_{1+1}(x)\delta b_{2}(y)}|_{J=b=0}.
\end{eqnarray}
By putting all these results together we obtain
\begin{eqnarray}
<e^{-\sigma(m,\tilde{M})}>&=&1\nonumber\\
&-&4u\int d^Dx <Tr\tilde{M}^2m^2(x)>_{\rm co}\nonumber\\
&-&2u\int d^Dx <Tr\tilde{M}m\tilde{M}m(x)>_{\rm co}\nonumber\\
&+&8u^2\int d^Dx\int d^Dy <Tr\tilde{M}m^3(x)Tr\tilde{M}m^3(y)>_{\rm co}\nonumber\\
&+&8u^2\bigg(\int d^Dx <Tr\tilde{M}^2m^2(x)>_{\rm co}\bigg)^2\nonumber\\
&+&2u^2\bigg(\int d^Dx <Tr\tilde{M}m\tilde{M}m(x)>_{\rm co}\bigg)^2\nonumber\\
&+&8u^2\bigg(\int d^Dx <Tr\tilde{M}^2m^2(x)>_{\rm co}\bigg)\bigg(\int d^Dy <Tr\tilde{M}m\tilde{M}m(y)>_{\rm co}\bigg)\nonumber\\
&+&8u^2\int d^Dx\int d^Dy <\tilde{M}^2m^2(x)Tr\tilde{M}^2m^2(y)>_{\rm co}\nonumber\\
&+&2u^2\int d^Dx\int d^Dy<Tr\tilde{M}m\tilde{M}m(x)Tr\tilde{M}m\tilde{M}m(y)>_{\rm co}\nonumber\\
&+&8u^2\int d^Dx \int d^Dy<Tr\tilde{M}^2m^2(x)Tr\tilde{M}m\tilde{M}m(y)>_{\rm co}\nonumber\\
&-&32 u^3\int d^Dx\int d^Dy <Tr\tilde{M}m^3(x)\tilde{M}m^3(y)>_{\rm co}\int d^Dz <Tr\tilde{M}^2m^2(z)>_{\rm co}\nonumber\\
&-&16 u^3 \int d^Dx\int d^Dy <Tr\tilde{M}m^3(x)\tilde{M}m^3(y)>_{\rm co}\int d^Dz <Tr\tilde{M}m\tilde{M}m(z)>_{\rm co}\nonumber\\
&-&32 u^3\int d^Dx\int d^Dy \int d^D z <Tr\tilde{M}m^3(x)\tilde{M}m^3(y)Tr\tilde{M}^2m^2(z)>_{\rm co}\nonumber\\
&-&16 u^3\int d^Dx\int d^Dy \int d^D z <Tr\tilde{M}m^3(x)\tilde{M}m^3(y)Tr\tilde{M}m\tilde{M}m(z)>_{\rm co}\nonumber\\
&+&\frac{32}{3} u^4\int d^Dx\int d^D y\int d^Dz\int d^Dw<Tr\tilde{M}m^3(x)Tr\tilde{M}m^3(y)Tr\tilde{M}m^3(z)Tr\tilde{M}m^3(w)>_{\rm co}\nonumber\\
&+&32 u^4 \bigg(\int d^Dx\int d^Dy <Tr\tilde{M}m^3(x)\tilde{M}m^3(y)>_{\rm co}\bigg)^2.
\end{eqnarray}
The non-perturbative correction to the effective action (up to the fourth power in the field $\tilde{M}$) is therefore given by \footnote{To go from this formula to the one used in the text we must still scale the fields as $\tilde{M},m\longrightarrow \tilde{M}/\sqrt{\nu_2}, m/\sqrt{\nu_2}$ and then scale the coupling as $u\longrightarrow u/N$. The definition of $u$ changes then as $u=\nu_2\lambda/4!\longrightarrow u=N\lambda/(4!\nu_2)$.}

\begin{eqnarray}
\Delta S_{\rm eff}[\tilde{M}]&=&4u\int d^Dx <Tr\tilde{M}^2m^2(x)>_{\rm co}+2u\int d^Dx <Tr\tilde{M}m\tilde{M}m(x)>_{\rm co}\nonumber\\
&-&8u^2\int d^Dx\int d^Dy <Tr\tilde{M}m^3(x)Tr\tilde{M}m^3(y)>_{\rm co}\nonumber\\
&-&8u^2\int d^Dx\int d^Dy <\tilde{M}^2m^2(x)Tr\tilde{M}^2m^2(y)>_{\rm co}\nonumber\\
&-&2u^2\int d^Dx\int d^Dy<Tr\tilde{M}m\tilde{M}m(x)Tr\tilde{M}m\tilde{M}m(y)>_{\rm co}\nonumber\\
&-&8u^2\int d^Dx \int d^Dy<Tr\tilde{M}^2m^2(x)Tr\tilde{M}m\tilde{M}m(y)>_{\rm co}\nonumber\\
&+&32 u^3\int d^Dx\int d^Dy \int d^D z <Tr\tilde{M}m^3(x)Tr\tilde{M}m^3(y)Tr\tilde{M}^2m^2(z)>_{\rm co}\nonumber\\
&+&16 u^3\int d^Dx\int d^Dy \int d^D z <Tr\tilde{M}m^3(x)\tilde{M}m^3(y)Tr\tilde{M}m\tilde{M}m(z)>_{\rm co}\nonumber\\
&-&\frac{32}{3} u^4\int d^Dx\int d^D y\int d^Dz\int d^Dw<Tr\tilde{M}m^3(x)Tr\tilde{M}m^3(y)Tr\tilde{M}m^3(z)Tr\tilde{M}m^3(w)>_{\rm co}.\label{primary}\nonumber\\
\end{eqnarray}
The four terms ($2$nd, $5$th, $6$th and the $8$th) containing the operator $Tr\tilde{M}m\tilde{M}m$ are subleading in the large $N$ limit. Each one of these diagrams is non-planar in the sense that if drawn on the plane then at least one of the $\tilde{M}$ fields will not be connected to the outside of the diagram, i.e. it will intersect .

\section{The Propagator}
We start with the Laplacian 
\begin{eqnarray}
G_{mn,kl}=\bigg(p^2+\mu^2+r^2(m+n-1)\bigg)\delta_{ml}\delta_{nk}+\pi r^2 \kappa^2\delta_{mn}\delta_{kl}.
\end{eqnarray}
The propagator is defined by
\begin{eqnarray}
\sum_{k,l}\Delta_{nm,lk}G_{kl,ts}=\sum_{k,l}G_{mn,kl}\Delta_{lk,st}=\delta_{ns}\delta_{mt}.
\end{eqnarray}
Since the Laplacian $G_{mn,kl}$ is non-zero only when $m+k=n+l$ the propagator $\Delta_{mn,kl}$ is non-zero only when  $m+k=n+l$. We introduce $n=m+\alpha$ and $s=t+\beta$ then
\begin{eqnarray}
\sum_{l}G_{mm+\alpha,l+\alpha l}\Delta_{ll+\alpha ,t+\beta t}=\delta_{\alpha \beta}\delta_{mt}.
\end{eqnarray}
We remark that for $\alpha\neq \beta$ we have immediately 
\begin{eqnarray}
\Delta_{ll+\alpha,t+\beta t}=0. \label{pro1}
\end{eqnarray}
For $\alpha=\beta$ we obtain an ordinary matrix inversion for every value of $\alpha$, viz
\begin{eqnarray}
\sum_{l}G_{mm+\alpha,l+\alpha l}\Delta_{ll+\alpha ,t+\alpha t}=\delta_{mt}.
\end{eqnarray}
We introduce  $G_{ml}^{\alpha}=G_{mm+\alpha,l+\alpha l}$ and $\Delta_{lt}^{\alpha}=\Delta_{ll+\alpha ,t+\alpha t}$ we write this as 
\begin{eqnarray}
\sum_{l}G_{ml}^{\alpha}\Delta_{lt}^{\alpha}=\delta_{mt}.
\end{eqnarray}
We have
\begin{eqnarray}
G_{ml}^{\alpha}=\bigg(p^2+\mu^2+r^2(2m+\alpha-1)\bigg)\delta_{ml}+\pi r^2\kappa^2\delta_{\alpha 0}.
\end{eqnarray}
Thus for $\alpha=\beta$ and $\alpha\neq 0$ we have immediately 
\begin{eqnarray}
\Delta_{ml}^{\alpha}=\frac{1}{p^2+\mu^2+r^2(2m+\alpha-1)}\delta_{ml}.\label{pro2}
\end{eqnarray}
The only  value which requires special attention is $\alpha =0$. The propagator $\Delta^0_{mt}$ satisfies the equation 
\begin{eqnarray}
\bigg(p^2+\mu^2+r^2(2m-1)\bigg)\Delta^0_{mt}+\pi r^2\kappa^2\sum_l\Delta^0_{lt}=\delta_{mt}.
\end{eqnarray}
This can be solved by the ansatz
 \begin{eqnarray}
\Delta^0_{mt}=\frac{1}{p^2+\mu^2+r^2(2m-1)}\bigg[\delta_{mt}+X_t\bigg].
\end{eqnarray}
We can find immediately 
 \begin{eqnarray}
X_t=-\frac{\pi r^2\kappa^2}{1+\pi r^2\kappa^2\sum_l\big[p^2+\mu^2+r^2(2l-1)\big]^{-1}}\frac{1}{p^2+\mu^2+r^2(2t-1)}.
\end{eqnarray}
In this paper we are dealing with the limit $\theta\longrightarrow \infty$ and hence we do not really need the above exact solution of  $\Delta^0_{mt}$. In fact all we need is the leading correction in the new parameter $r^2\kappa^2$ (recall that $r^2=4/\theta$). From the above solution we obtain in a straightforward way the result
 \begin{eqnarray}
\Delta^0_{mt}=\frac{1}{p^2+\mu^2+r^2(2m-1)}\bigg[\delta_{mt}-\frac{\pi r^2\kappa^2}{p^2+\mu^2+r^2(2t-1)}+O(r^4\kappa^4)\bigg].\label{pro3}
\end{eqnarray}
From equations (\ref{pro1}), (\ref{pro2}) and (\ref{pro3}) we get the propagator 
\begin{eqnarray}
\Delta_{mn,kl}=\frac{1}{p^2+\mu^2+r^2(m+n-1)}\bigg[\delta_{ml}\delta_{nk}-\frac{\pi r^2\kappa^2}{p^2+\mu^2+r^2(k+l-1)}\delta_{mn}\delta_{kl}+O(r^4\kappa^4)\bigg].\nonumber\\
\end{eqnarray}

\end{document}